\documentclass[letterpaper,twocolumn,10pt]{article}

\usepackage{usenix2019_v3}
\usepackage[utf8]{inputenc}
\usepackage[T1]{fontenc}
\usepackage{microtype}
\usepackage{booktabs}
\usepackage{tabularx}
\usepackage{multirow}
\usepackage{graphicx}
\usepackage{xcolor}
\usepackage{amsmath}
\usepackage{amssymb}
\usepackage{hyperref}
\usepackage{cleveref}
\usepackage{enumitem}
\usepackage{xspace}
\usepackage{balance}
\usepackage{listings}
\usepackage{pifont}

\lstdefinestyle{cpp}{
  basicstyle=\ttfamily\footnotesize,
  keywordstyle=\bfseries,
  commentstyle=\itshape\color{gray},
  language=C++,
  columns=fullflexible,
  showstringspaces=false,
  breaklines=true,
  frame=single,
  framesep=3pt,
  xleftmargin=4pt,
  aboveskip=4pt,
  belowskip=4pt,
}

\hypersetup{
  colorlinks=true,
  linkcolor=blue!60!black,
  citecolor=blue!60!black,
  urlcolor=blue!60!black,
}

\setlist{nosep,leftmargin=1.2em}

\newcommand{\visp}{\textsc{VISP}\xspace}
\newcommand{\oc}{\textsc{OC}\xspace}
\newcommand{\skill}{\textsc{SKILL}\xspace}

\title{Mixed-Initiative Mining of Root-Cause Analyses into\\
Audit Priors for C/C++ Vulnerability Hunting}

\author{
  Yuhei Kawakoya \qquad Eitaro Shioji \qquad Yuto Otsuki \qquad Makoto Iwamura \\[4pt]
  NTT Social Informatics Laboratories
}

\date{}

\begin{document}
\maketitle

\begin{abstract}
Vulnerability auditing of large C/C++ codebases is constrained by attention:
human auditors and LLM-based agents cannot inspect every file and function
uniformly. Budgeted auditing instead requires reusable \emph{operational
priors} answering two questions---\emph{where} limited review effort should be
spent, and \emph{what} should be suspected once a location is selected---yet
such priors rarely exist in a form that an auditor or agent can directly
consume. This paper does \emph{not} propose a new vulnerability detector; it
proposes RCA-derived audit-steering artifacts that supply these priors.

We mine evidence-backed root-cause analysis (RCA) reports of past C/C++
vulnerabilities into two complementary artifacts. Occurrence Context (\oc{}) is
a location prior: a three-axis schema (file role / function role /
data-structure class) that specifies \emph{where to look first}.
Vulnerability-Inducing Structural Patterns (\visp{}) are failure-mechanism
priors: reusable structural patterns that specify \emph{what to suspect there}.

Evaluated on 108 CVEs from Vul4C, \oc{}'s parser-based, no-LLM priority
ranking outperforms generic prioritization baselines with moderate-to-large
effect sizes. Packaged as executable \skill{}s, the combined dual-\skill{}
workflow improves audit-relevant 3-LLM-judge broad coverage---match-or-partial
findings inside the ground-truth function that share the patch's
failure mechanism. At the aggregate level
dual-\skill{} adds $+5.3$\,pp Func-level broad matches across the
108-CVE cohort, and this gain is scale-conditioned, rising to
$+7.2$\,pp on mid-size projects and $+15.0$\,pp on large
(1500--4999 line) vulnerable files. Overall, RCA-derived SKILLs act as
audit-steering knowledge: they broaden where an auditor's limited
attention lands inside ground-truth vulnerable code. We report this
agent-coverage effect as a single-run operational probe---measured
once per commit, characterizing sign and scale rather than a precise
effect size.

\end{abstract}

\begin{figure*}[t]
\centering
\includegraphics[width=0.95\textwidth]{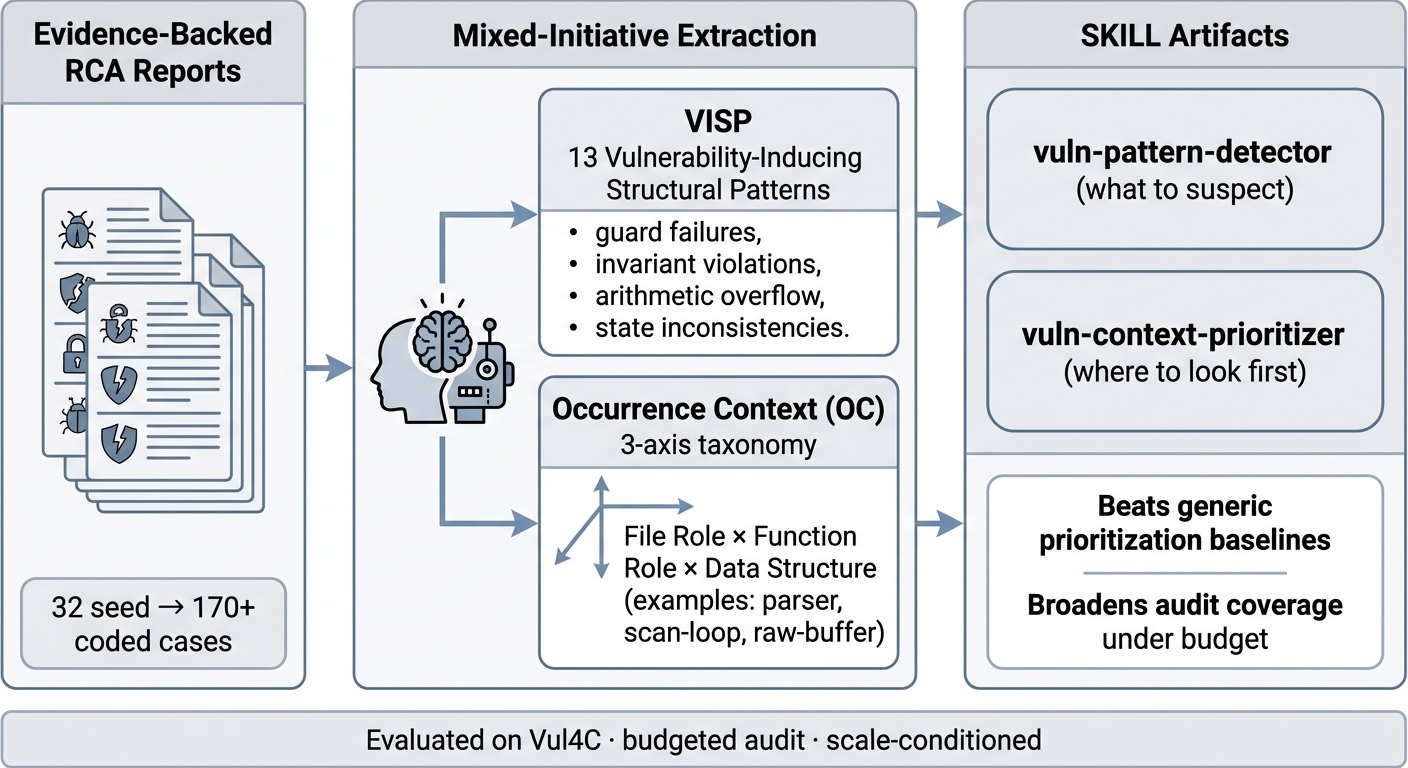}
\caption{Graphical abstract. Evidence-backed RCA reports are mined through a
mixed-initiative extraction process into two complementary intermediate representations
(\visp{} and \oc{}), which are then packaged as reusable \skill{}
artifacts for operational vulnerability hunting.}
\label{fig:graphical_abstract}
\end{figure*}

\section{Introduction}
\label{sec:intro}

C/C++ vulnerabilities in large codebases rarely arise as isolated events:
they recur around a bounded set of program structures, metadata-handling
routines, and buffer-management idioms, and experienced reviewers exploit
this regularity implicitly rather than reading the repository uniformly.
Auditing such repositories is therefore as much an
\emph{attention-allocation} problem as a detection problem. Unlike static
analyzers, which can be applied mechanically across whole codebases, human
reviewers and LLM-based agents operate under limited reading budgets. Two
complementary priors then govern what the audit surfaces: one says
\emph{where to look} first, and the other says \emph{what to suspect} once
there---and neither is directly supplied by existing vulnerability
knowledge sources.

The difficulty is that this expert knowledge is not usually available in an
operational form. High-level taxonomies such as CWE~\cite{MITRE_CWE} are useful
for reporting and aggregation, but they are too coarse to directly steer
concrete code reading. Patch-pattern mining emphasizes how bugs were fixed
rather than what vulnerable structure existed before the
fix~\cite{Kim_2013,Bader_2019}. Prescriptive secure-coding standards such as
SEI CERT C~\cite{Seacord_2014} provide critical normative guidance, but they
primarily express obligations over language elements and local checks, not the
recurring,
empirical structures that appear across vulnerability cases.

Evidence-backed root-cause analysis (RCA) reports sit between raw patches and
coarse taxonomies. A patch records \emph{how} a vulnerability was fixed, and
a CWE label records \emph{how} the failure is classified, but an RCA explains
\emph{why} the pre-fix structure was vulnerable and \emph{where} that
structure occurred. Mining RCAs therefore exposes two audit-relevant priors
that are difficult to recover from patches or CWE labels alone: a
location-oriented prior for prioritizing files and functions, and a
mechanism-oriented prior for structuring suspicion inside the selected code.

This work extracts reusable audit-steering artifacts from evidence-backed
root-cause analysis (RCA) reports of past C/C++ vulnerabilities, so that
both human reviewers conducting manual code reading and LLM-based agents
performing automated scanning can use the same operational priors, and
so that the artifacts are operationalizable as executable workflows
rather than a descriptive taxonomy. We use \skill{} to denote an
agent-loadable bundle of instructions, reference tables, and output
schemas; conceptually, the artifacts are agent-readable audit playbooks
rather than a detector tied to a particular agent framework.

We produce two complementary representations. The first,
Vulnerability-Inducing Structural Patterns (\visp{}), captures recurring
program structures that induce vulnerabilities---for example, access
without a valid guard at the use site, cross-field consistency failures
between header-derived values, or arithmetic overflow before a bound
check---as named patterns with tiered structural cues. The second,
Occurrence Context (\oc{}), models \emph{where} vulnerabilities tend to
cluster through a three-axis schema of file role (e.g., parser,
decoder), function role (e.g., scan-loop, header-init), and
data-structure class (e.g., raw-buffer, header-struct). Together,
\visp{} tells an auditor what to suspect (a failure-mechanism prior)
and \oc{} tells them where to look first (a location prior),
restructuring the order in which code is explored and broadening the
set of plausible findings returned, rather than improving pinpoint
localization of any individual finding.

We organize the evaluation around three research questions, with weight
asymmetric by design: RQ2 is the paper's hardest evidence, RQ3 is the
operational claim, and RQ1 is a prerequisite rather than a contribution
in its own right.
\textbf{RQ1 (Representation Stability)} asks whether evidence-backed
RCA reports can be mined into stable, reusable intermediate
representations that remain coherent when the corpus grows beyond the
seed cases. \textbf{RQ2 (Priority Signal Quality)} asks whether \oc{}'s
context-based priority ranking, isolated from \visp{}, outperforms
generic prioritization baselines---random selection, cyclomatic
complexity Top-$N$~\cite{McCabe_1976}, and SAST alert density---under a
matched function-budget. \textbf{RQ3 (Budgeted Audit Coverage)} asks
whether the combined dual-\skill{} workflow, applied as an
LLM-agent-consumed policy, improves audit-relevant broad-match
detection inside ground-truth functions---findings that share the
patch's failure mechanism, the criterion auditors likely use---
and how the effect scales
with project size and vulnerable-file size. A
replay-style analysis of the mining process is reported in
\Cref{app:replayability}.

The paper makes three contributions. First, it introduces two
RCA-derived audit-steering artifacts---\visp{} for structural mechanisms
(\emph{what to suspect}) and \oc{} for location-sensitive audit priors
(\emph{where to look})---and packages them as executable
\skill{}s. Second, the \emph{operationalized} \oc{} ranker---a
parser-based, no-LLM ranker derived from the \oc{} taxonomy---outperforms
random selection, cyclomatic complexity Top-$N$, and SAST alert density
on 108 CVEs with moderate-to-large effect sizes
($d=0.60$--$1.03$); this is the paper's hardest result. Third, the
dual-\skill{} workflow improves the audit-relevant 3-LLM-judge broad-match
inside ground-truth functions---findings that share the patch's
failure mechanism even if they target a related defect or a different
facet of a multi-location fix---by $+5.3$\,pp at the Func granularity
and $+8.6$\,pp at the Line~$\pm 30$ granularity at the aggregate
level under 3-LLM-judge triangulation; the gain is scale-conditioned
(\Cref{sec:rq3}).

Beyond these results, we document the governed mixed-initiative
protocol used to produce \visp{} and \oc{}---human decision gates,
update rules, and a 77-run replay-style ablation showing AI
reliably replays the research pathway while human commitment fixes
the concrete artifact---so the artifacts are reproducible as a
process, not only as outputs (\Cref{sec:extraction},
\Cref{app:rules}, \Cref{app:replayability}).

\section{Background and Related Work}
\label{sec:related}

Vulnerability taxonomies such as CWE~\cite{MITRE_CWE,CWE_Top25} are
indispensable for aggregation, reporting, and communication, yet they generally
operate at a level of abstraction too coarse to guide repository-scale auditing
directly. A category such as buffer over-read, use-after-free, or integer
overflow describes the outward failure class, but it does not necessarily tell a
reviewer which structural configuration in code should be suspected during
manual or LLM-assisted reading. Recent efforts have extended CWE with more
specific weakness chains and composite descriptions~\cite{CWE_Chains}, yet the
gap between taxonomy-level labels and concrete code-reading guidance persists.

A substantial body of work extracts patterns from vulnerability-fixing commits,
security patches, and bug
databases~\cite{Kim_2013,Le_2016,Bader_2019,LeGoues_2019}. Patch-template
mining learns fix patterns that can be reapplied to similar bugs, but the
templates emphasize the transformation rather than the prior vulnerable
structure. Bug-pattern detection identifies recurring code-level
anti-patterns~\cite{Hovemeyer_2004,Yamaguchi_2014}, but often at the syntactic
level without the semantic context of why the structure is dangerous in a
particular functional role. Vulnerability prediction models use features
extracted from code or metadata to rank files or functions by
risk~\cite{Shin_2011,Morrison_2015,Li_2022}, but they typically produce opaque
scores rather than interpretable structural explanations. Static analysis tools
have demonstrated the value of systematic code examination at
scale~\cite{Bessey_2010}, while deep-learning- and LLM-based
approaches have broadened the scope of automated vulnerability
handling---from learned-feature detection~\cite{Li_2018} to
LLM-driven repair~\cite{Pearce_2023} and autonomous
exploitation~\cite{Fang_2024}. Our approach differs in that it
mines pre-fix vulnerable structures from RCA reports rather than post-fix
transformations, and it produces named, interpretable patterns that an
analyst can use during code reading.

Prescriptive secure-coding guidelines such as SEI CERT
C~\cite{Seacord_2014,CERTC_Wiki} provide normative guidance that is critical for
secure development and review. Each rule
addresses specific language-element concerns such as integer overflow, null
pointer dereference, or string handling. These rules provide necessary
discipline, but their organization around language elements and local checks
makes it difficult to express recurring cross-field, cross-component, or
parser-specific failure structures that appear in real postmortems. Our
comparison in \Cref{sec:representations} and \Cref{app:visp} shows that
\visp{} occupies a complementary space: nine of thirteen \visp{}
patterns map directly to existing CERT~C rule families, one is only
weakly aligned, and three capture structural failure modes that have
no directly corresponding rule.

Emerging work on human--AI collaboration in research has explored co-authoring,
hypothesis generation, and data-driven
analysis~\cite{Wang_2023,Wu_2022}. The methodological question most relevant to
our work is not merely whether AI can help extract patterns, but which parts of
the research process remain distinctly human when AI can also replay candidate
generation, comparison, and schema construction. Our staged ablation study
provides empirical evidence that the research pathway is reliably replayable by
AI, while the research product depends on human commitment, validation
investment, and completion judgment---offering a more precise framework than the
common framing of ``human creativity plus AI scale.''

\Cref{tab:positioning} summarizes the positioning of our work relative to these
four related streams.

\begin{table*}[t]
\centering
\caption{Positioning of this work relative to related streams.}
\label{tab:positioning}
\small
\begin{tabular}{@{}lllll@{}}
\toprule
\textbf{Dimension} & \textbf{Taxonomy (CWE)} & \textbf{Patch mining} & \textbf{Guidelines (CERT C)} & \textbf{This work (\visp{} + OC)} \\
\midrule
Primary input & Reported vulns & Fix commits & Language spec & Evidence-backed RCA reports \\
Abstraction level & Failure class & Repair template & Lang-element rule & Structural mechanism + context \\
Operational target & Reporting & Automated repair & Dev compliance & Code reading \& triage \\
Interpretability & High (taxonomy) & Medium (template) & High (rule) & High (pattern + schema) \\
Location guidance & None & Implicit (patch site) & Implicit (rule scope) & Explicit (\oc{}) \\
\bottomrule
\end{tabular}
\end{table*}

\section{Study Design and RCA Substrate}
\label{sec:design}

A key design decision of this study is to separate the dataset used to
\emph{mine} the representations from the dataset used to \emph{evaluate}
them. The two datasets have distinct origins and are independently curated,
so that the RQ2 and RQ3 evaluations are genuinely held-out tests rather
than in-sample replays.

\paragraph{Mining substrate (PatchAgent dataset).}
The starting point for mining \visp{} and \oc{} is the dataset released
with \textsc{PatchAgent}~\cite{Yu_2025}, a USENIX Security~2025
program-repair system. The \textsc{PatchAgent} dataset contains 178
real-world vulnerabilities across \emph{30 original C/C++ programs}
(28 cases from the ExtractFix corpus and 150 cases from OSS-Fuzz and
Huntr), covering nine bug types including heap/stack/global/integer
overflows, use-after-free, double-free, invalid-free, null
dereference, and divide-by-zero. The shipped dataset organizes these
178 cases into \emph{33 project contexts}: the 30 original programs
plus 3 additional build contexts for parser-heavy targets in which
the same upstream code is exercised under a different build/harness
configuration. The 3 additions are an ExtractFix-corpus libxml2
build (\texttt{extractfix-libxml2}), an ExtractFix-corpus libxml2
fuzz-harness build (\texttt{extractfix-libxml2-harness}), and a
wasm3 fuzz-harness build (\texttt{wasm3-harness}); the corresponding
upstream programs (\texttt{libxml2} and \texttt{wasm3}) are also
present as their own contexts. We carry this 33-context layout
forward because each context exposes distinct file/function
boundaries to the analysis pipeline.

\paragraph{RCA generation and the case unit.}
The shipped \textsc{PatchAgent} dataset provides patch hunks, affected
code regions, vulnerability metadata, and a proof-of-crash (PoC) input,
but not narrative root-cause analyses. We therefore generate the RCA
reports ourselves with an evidence-backed pipeline: for each case we
build the vulnerable binary under AddressSanitizer, reproduce the crash
with the shipped PoC, and an LLM agent then synthesizes the patch hunks,
the surrounding code, the metadata, the resulting sanitizer report and
crash stack, and GDB sessions (breakpoints, watchpoints, variable
inspection) into an RCA narrative tying the pre-fix structure, the
failure mechanism, and the observed effect to specific code locations.
Every claim is cited to one of these evidence sources. These RCA
reports---grounded in explicit static and dynamic crash evidence rather
than free-form intuition---are the substrate from which \visp{} and
\oc{} are mined. We refer to each
\textsc{PatchAgent} entry---one vulnerability instance with patch
hunks, affected code, and our generated RCA narrative---as a
\emph{case}, since not every entry carries a CVE identifier (some
originate from OSS-Fuzz or Huntr without a CVE assignment); we reserve
the term \emph{CVE} for entries identified by a CVE number, used in
the downstream evaluation.

\paragraph{Seed and incremental coverage.}
From this substrate we sample one case per project context as the
seed set, targeting 33 cases. One context's seed case
(\texttt{coreutils}) timed out during RCA generation, leaving
\emph{32 seed cases spanning 32 of the 33 project contexts}. We then
extend incrementally through five batches to a final corpus of 173
coded cases for \visp{} and 172 for \oc{} (\Cref{tab:scale_up}). The
missing context (\texttt{coreutils}) was recovered in the first
incremental batch through three of its other cases, so the mining
corpus spans all 33 \textsc{PatchAgent} project contexts
(equivalently, all 30 original C/C++ programs), including
parser-heavy targets such as GPAC, radare2, libxml2, hunspell, mruby,
and wasm3.

\paragraph{Downstream evaluation (Vul4C).}
For the operational evaluations (RQ2 and RQ3) we use the Vul4C
benchmark~\cite{Hu_2025}, a separate public dataset of 144~CVEs
across 23 C/C++ projects, originally designed for automated
program-repair evaluation. We retain a \emph{108-CVE cohort across
19 C/C++ projects and 51 unique fix commits} that we use throughout
RQ2 and RQ3. The 36 dropped CVEs fall into four non-evaluative
buckets ($10 + 9 + 3 + 14 = 36$): (i)~mining-corpus overlap
(held-out integrity), (ii)~upstream metadata gaps, (iii)~build-pipeline
format-conversion losses, and (iv)~CVEs that fall outside the RQ2
prioritization scope (Buffer, Integer, UAF, NPD, Input-Validation
families). A complete CVE-level audit trail with
per-bucket listings, the per-project breakdown of the 108-CVE
cohort, and the reduction chain
$144\to134\to125\to122\to108$ is given in \Cref{app:cohort_audit}.

This separation between mining substrate and evaluation benchmark lets us
study four distinct phenomena within one paper: whether the representations
remain stable under corpus growth (RQ1), whether \oc{}'s priority signal
beats generic baselines (RQ2), whether the combined dual-\skill{} workflow
is operationally useful at benchmark scale (RQ3), and whether the human
role can be isolated through replay-style ablations
(\Cref{app:replayability}).

\section{Mixed-Initiative Knowledge Extraction}
\label{sec:extraction}

\begin{figure*}[t]
\centering
\includegraphics[width=0.95\textwidth]{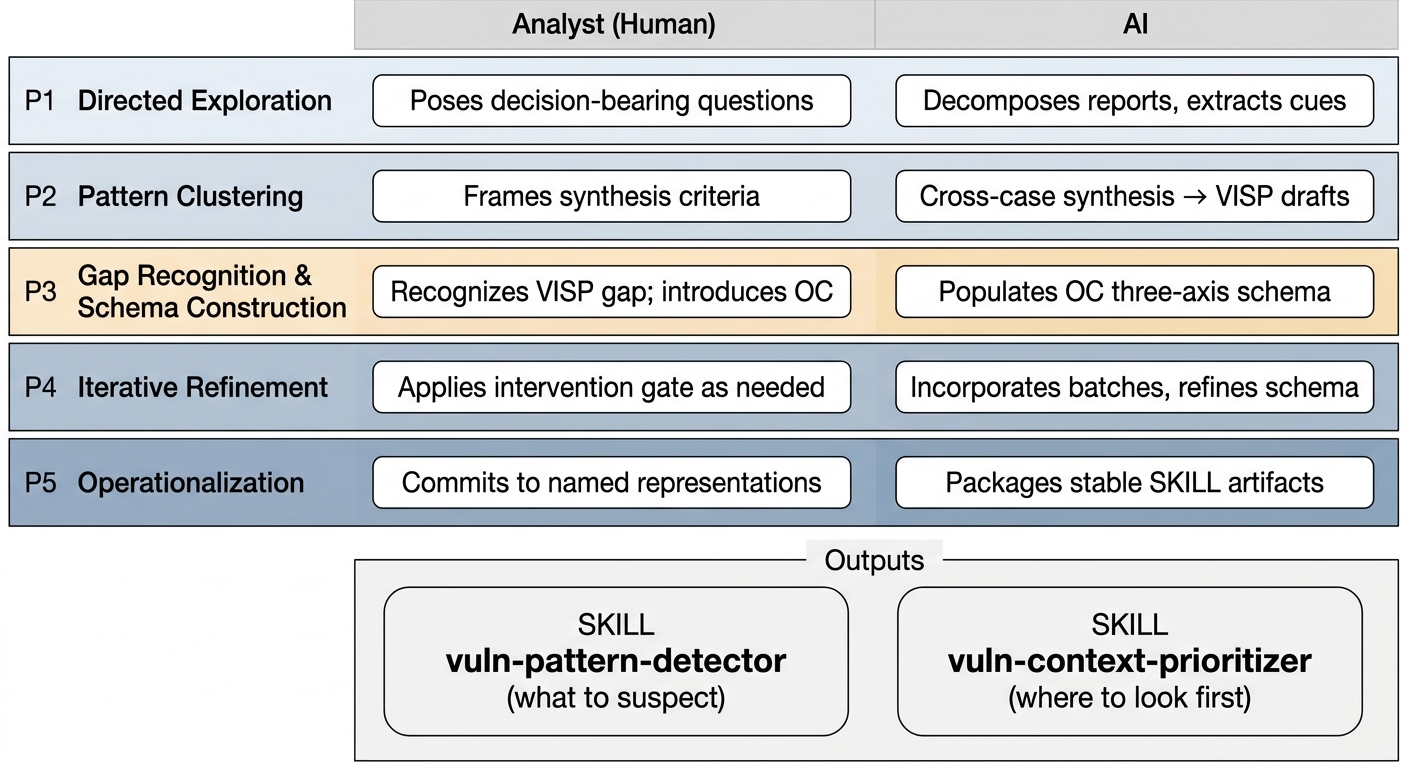}
\caption{Mixed-initiative knowledge extraction process. Columns
separate Analyst (left) and AI (right) contributions; rows trace
five phases (P1--P5) that run top-to-bottom. P1--P3 form the
intermediate representations \visp{} and \oc{} from the seed RCA
substrate; P4--P5 stabilize them and package the two \skill{}
artifacts (\texttt{vuln-pattern-detector} and
\texttt{vuln-context-prioritizer}). Each phase is detailed in
\Cref{sec:extraction}.}
\label{fig:extraction}
\end{figure*}

This section describes the governed mixed-initiative protocol used
to convert evidence-backed RCA reports into committed, reusable
audit-steering artifacts. The goal is to characterize the controlled
human-AI workflow through which RCA evidence was converted into
\visp{} and \oc{}: how human framing, AI-assisted synthesis, human
commitment, and incremental validation interact to produce the two
representations. The central security claims are
evaluated later---\oc{}'s deterministic prioritization signal in RQ2
(\Cref{sec:rq2}) and the dual-\skill{} audit-coverage effect in RQ3
(\Cref{sec:rq3}). The role of this section is to explain why the
resulting \visp{} and \oc{} artifacts are not ad-hoc prompt
products: they are committed representations produced under human
framing, AI-assisted synthesis, explicit promotion / update rules
(\Cref{app:rules}), and incremental validation, with a recorded
audit trail of human contributions and AI difficulties
(\Cref{app:human_ai}).

The process proceeds through five phases (P1--P5), as illustrated
in \Cref{fig:extraction}. The first three phases form intermediate
representations from the seed RCA substrate: \visp{} is formed
through P1 (Directed Exploration) and P2 (Pattern Clustering)
under a directed question about recurring vulnerability-inducing
program structures (\Cref{subsec:visp_formation}), and \oc{} is
admitted in P3 (Gap Recognition and Schema Construction) after
the analyst recognizes that \visp{} alone cannot express the
location-sensitive priors needed for budgeted auditing
(\Cref{subsec:oc_repair}). The remaining two phases stabilize
both representations: P4 (Iterative Refinement) absorbs additional
RCA batches under a human decision gate, and P5
(Operationalization) packages the refined representations as the
two \skill{} artifacts---both governed by the explicit promotion,
update, and stopping rules in \Cref{subsec:governance}. The
protocol is governed in the sense that the AI proposes,
decomposes, clusters, and populates, while the human frames the
question, admits each representation, commits names and
boundaries, decides when to stop, and turns the output into
reusable \skill{} artifacts.

\subsection{Directed \visp{} Formation}
\label{subsec:visp_formation}
\visp{} was the result of a directed search. The analyst framed the
seed-stage question around program structures that tend to induce
vulnerabilities, rather than around repair templates or CWE
relisting, and asked the AI to synthesize recurring structural
mechanisms across cases. Under that framing, the AI
repeatedly converged on representations centered on guard failures,
invariant violations, state inconsistencies, and arithmetic-to-memory
propagation. These families were committed as named patterns by the
analyst because they could be grounded in RCA evidence, reused across
cases, and later operationalized into a review artifact. The
analyst, not the AI, decided which candidate families to admit as
first-class representations. What is replayable here is therefore
the directed pathway from seed RCA evidence to a \visp{}-equivalent
representation, given the same framing and quality gates---not the
specific transcript that produced any one pattern.

\subsection{\oc{} as Representational Repair}
\label{subsec:oc_repair}
\oc{} entered the process for a different reason. Once \visp{} made
structural mechanisms more legible, the analyst recognized a
practical shortfall in audit workflow terms: \visp{} answers
\emph{what to suspect}, but it does not tell an auditor
\emph{where to look first} under a budget. The analyst identified
this as an audit-workflow gap and used iterative AI-assisted schema
refinement to turn it into the \oc{} representation, which captures
file role, function role, and data-structure class. \oc{}
does not replace \visp{}; the two are complementary because they
target different bottlenecks. \oc{} addresses the
repository-level selection bottleneck (which files / functions to
read first), and \visp{} addresses the within-file focus bottleneck
(what to suspect once a function is being read). Both interpretations
are tested in the operational evaluations: \oc{} is isolated under
RQ2 (\Cref{sec:rq2}) as a deterministic prioritization signal, and
the joint effect of the two as audit priors is the subject of RQ3
(\Cref{sec:rq3}).

\subsection{Governance, Update, and Stopping Rules}
\label{subsec:governance}
The protocol uses three explicit gates, recorded in
\Cref{app:rules}, to keep the catalog from drifting under scale-up.
First, a candidate abstraction is promoted to a first-class
representation only if it (i)~can be filled in from the RCA evidence,
(ii)~answers an audit question that existing representations do not
already cover, (iii)~has stated definitions and boundary conditions,
and (iv)~is expected to remain stable under bounded refinement when
the corpus grows. \visp{}'s 13 patterns and the three \oc{} axes
were each admitted under this rule.

Second, during incremental coding (P4), new RCA cases are absorbed
into existing labels by default; new labels are introduced only when
repeated semantic mismatch makes absorption infeasible. Every
merge / split / add decision is recorded with a rationale so that
the catalog's growth can be replayed against the same evidence.

Third, stopping is a human decision, not an AI one. The analyst
declares the catalog stable when bounded refinement persists across a
batch under both rules above; the resulting stability is later
quantified as RQ1. \Cref{app:human_ai} records cases where AI
synthesis exhibited tendencies---most prominently no self-imposed
stopping criterion and a tendency toward over-fragmentation when
schema contracts were absent---that the human gates were specifically
designed to absorb.

\section{Intermediate Representations and Artifactization}
\label{sec:representations}

The mining process (\Cref{sec:extraction}) distills two complementary
representations---\visp{} for recurring structural mechanisms
(\emph{what to suspect}) and \oc{} for location-sensitive triage priors
(\emph{where to look})---together with a compact scoring procedure that
turns \oc{}'s labels into a priority tier. Below we describe each
representation and the executable \skill{} artifacts that package them
for agent consumption.

\paragraph{\visp{}: vulnerability-inducing structural patterns.}
\visp{} captures recurring vulnerability-inducing program structures as
reusable pattern entries. Each entry specifies the mechanism, detection
cues, expected RCA evidence, typical CWE associations, and grounding cases
used to calibrate interpretation---a structured abstraction that an auditor
can use to recognize \emph{analogous} structures in new code rather than
match an exact code template. Detection cues are organized into two tiers: Tier-1 cues are
quick-scan signals visible during code skimming, while Tier-2 checks are deeper
confirmations performed after a Tier-1 cue is found. In the current catalog,
\visp{} spans 13 patterns (VISP-1 through VISP-13) categorized over
170 of 173 incrementally coded RCA narratives (three narratives did
not match any refined pattern definition; see \Cref{tab:visp_full}).
The seed catalog contained VISP-1 through VISP-10, and incremental
coding added three patterns: VISP-11 (unvalidated use of
external-input fields), VISP-12 (logical or variable mistakes inside
guard conditions), and VISP-13 (missing cycle detection in recursive
traversal). All three additions appeared within the first two
incremental batches, after which the catalog stabilized. The top
five patterns alone account for 65\% of all categorized cases,
meaning that a small number of structural suspicion templates can
cover the majority of the vulnerability space. \Cref{tab:visp_excerpt} shows representative patterns.

\begin{table}[t]
\centering
\caption{Representative \visp{} patterns (excerpt).}
\label{tab:visp_excerpt}
\small
\begin{tabular}{@{}lrp{3.2cm}@{}}
\toprule
\textbf{Pattern} & \textbf{Cases} & \textbf{Core mechanism} \\
\midrule
VISP-1  & 33 & Access without valid guard at use site \\
VISP-4  & 16 & Cross-field consistency not enforced \\
VISP-8  & 22 & Computed limit $\neq$ actual accessed range \\
VISP-11 & 17 & External field used in dangerous operation \\
\bottomrule
\end{tabular}
\end{table}

\visp{} is structurally complementary to prescriptive secure-coding
guidelines: nine patterns map directly to SEI CERT~C~\cite{Seacord_2014}
rule families, one is weakly aligned, and three (VISP-4, 7, 10) have no
directly corresponding rule because they require cross-field invariant
reasoning, asymmetric trust-boundary analysis, or nested-boundary
parsing checks that exceed local rule scope. The full pattern-by-pattern
alignment and a worked VISP-4 example are in \Cref{app:visp}.

\paragraph{\oc{}: three-axis occurrence-context taxonomy.}
\oc{} models where vulnerabilities are likely to occur through a
compact three-axis schema: file role, function role, and data-structure class.
In the final taxonomy over 172 coded RCA results, \oc{} contains
10~file-role labels, 13~function-role labels, and 16~data-structure
labels (full label lists, the TOP15 high-priority combinations, and
the module-type-to-expected-\visp{} mapping are released in the
supplementary materials).
Compared with the seed taxonomy (10/12/16), the enlarged corpus adds only one
new label---\texttt{buffer-resize}---to distinguish dedicated growth/shrink
routines from generic parsing or memory-management roles. Thus, 38 of the final
39 labels (97.4\%) were already present in the seed set, and no further labels
were added after the 121st coded case. The final distribution is strongly
concentrated: \emph{parser} (42.4\%), \emph{scan-loop} (36.0\%), and
\emph{raw-buffer} (41.3\%) dominate their respective axes.
\Cref{tab:top5_context} shows the top five high-priority combinations, which
alone cover 36.0\% of all coded cases.

\begin{table*}[t]
\centering
\caption{Top 5 high-priority context combinations (covering 36.0\% of 172 coded cases).}
\label{tab:top5_context}
\footnotesize
\setlength{\tabcolsep}{4pt}
\begin{tabular}{@{}clrcr@{\hskip 6pt}l@{\hskip 6pt}l@{}}
\toprule
\textbf{\#} & \textbf{Context combination} & \textbf{Cases} & \textbf{Cum.} & \textbf{\visp{}} & \textbf{CWE} & \textbf{Risk mechanism} \\
\midrule
1 & parser / header-init / header-struct   & 17 & 9.9\%  & 4, 11   & 119, 20  & Untrusted header fields flow into alloc/index \\
2 & parser / scan-loop / raw-buffer        & 16 & 19.2\% & 1, 2, 8 & 125, 787 & Loop reads past buffer w/o bound check \\
3 & parser / header-init / section-header  & 11 & 25.6\% & 1, 2, 4 & 190, 119 & Arithmetic on section sizes overflows \\
4 & parser / scan-loop / dynamic-buffer    &  9 & 30.8\% & 1, 6    & 787, 416 & Realloc invalidates in-flight pointers \\
5 & decoder / scan-loop / raw-buffer       &  9 & 36.0\% & 1, 8    & 125, 119 & Decoder loop trusts external length field \\
\bottomrule
\end{tabular}
\end{table*}

\paragraph{Priority scoring and AST heuristics.}
Beyond the three-axis labels and the TOP15 table, \oc{} includes a
compact scoring procedure that turns per-function labels into a
priority tier that downstream reviewers can consume. Each function $f$
receives an integer score
$s(f) = \alpha \cdot r_{\text{fn}}(f) + \beta \cdot r_{\text{file}}(f)
+ \gamma \cdot r_{\text{ds}}(f) + \delta_{\text{TOP15}}(f)
+ \eta_1 \cdot [H_1(f)] + \eta_2 \cdot [H_2(f)]$,
where $r_{\text{fn}}$, $r_{\text{file}}$, $r_{\text{ds}}$ are
role-specific risk scores attached to the function-role, file-role,
and data-structure-class axis labels;
$\delta_{\text{TOP15}}$ is a fixed bonus ($+2$) awarded when the
function's three-axis tuple matches an entry in the TOP15 high-risk
combinations (\Cref{tab:top5_context}); and $H_1$, $H_2$ are two
AST-pattern heuristics that fire on structural cues empirically
associated with recurring vulnerability mechanisms.
$H_1$ (\emph{scan-loop over manually-advanced index}) fires on a
loop whose termination depends on a manually advanced integer index or
pointer into an externally-sized buffer without an explicit bound check
in the loop header---the structural shape behind most of our observed
out-of-bounds read cases.
$H_2$ (\emph{header-init to field assignment}) fires on an
initialization or dispatch routine that assigns externally-derived
header fields directly into downstream structures consumed as lengths,
counts, or offsets.
Functions are binned by score band into five descending-priority tiers
$P_1$--$P_5$; $P_1$ captures the highest-scoring functions (all four
signal groups co-firing strongly), $P_5$ captures functions that match
only the weakest axis alone, and the operationally relevant tier in
our evaluation is the cumulative set $P_1\!+\!P_2$, which spans
roughly 22\% of the function scope per CVE on average. Full weight
values and tier thresholds are provided in \Cref{app:oc_eval}.

\paragraph{Skill artifacts and dual-\skill{} workflow.}
To make the two representations consumable by LLM agents without
custom integration code, we package them as separate executable
\skill{} artifacts (Claude Code Agent Skills---self-contained
bundles of instructions, reference tables, and output schemas
loaded on demand) so that each can be reused or evaluated in
isolation: \texttt{vuln-pattern-detector} operationalizes \visp{}
(\emph{what to suspect}), and \texttt{vuln-context-prioritizer}
operationalizes \oc{} (\emph{where to look first}). The two
compose into a two-step \emph{dual-\skill{} workflow}---
referenced throughout the rest of the paper---in which \oc{}
first assigns file/function/data-structure roles and orders the
review queue by tier $P_k(f)$ from the priority score $s(f)$,
then \visp{} is applied in descending priority order with
Tier-1 patterns as quick-scan filters and Tier-2 patterns as
deeper confirmations wherever a Tier-1 cue or a context-expected
pattern fires. Keeping the two artifacts separate enables testing
of context-only, pattern-only, and dual-composition modes.

\section{Evaluation}
\label{sec:eval}

The evaluation is organized around three research questions
(\Cref{tab:eval_roadmap}), weighted asymmetrically by design: RQ2
carries the paper's hardest evidence (parser-based, deterministic
ranker, no LLM in the loop at ranking time), RQ3 reports the
operational consequence under its claim boundary (\Cref{sec:rq3}),
and RQ1 is a prerequisite. A complementary replayability ablation
of the mixed-initiative mining process is reported in
\Cref{app:replayability}. All agent workflows use Anthropic's
Claude family at default temperature (Claude Opus 4.6
vulnerability scanner; Claude Sonnet 4.6 source-retrieval and
report-generation); LLM-as-judge adjudication for the three-judge
triangulation in \Cref{sec:rq3} uses Claude Opus 4.7, ChatGPT 5.4
Thinking, and Gemini 2.5 Pro.

\begin{table*}[t]
\centering
\caption{Evaluation roadmap.}
\label{tab:eval_roadmap}
\footnotesize
\setlength{\tabcolsep}{4pt}
\begin{tabular}{@{}lp{5.4cm}rp{5.6cm}@{}}
\toprule
\textbf{RQ} & \textbf{Core claim} & \textbf{$N$} & \textbf{Key metrics} \\
\midrule
RQ1 & \visp{} and \oc{} are stable, reusable representations & 173/172 coded & Growth rate, Chao1 saturation, within-coder Fleiss $\kappa$ \\
RQ2 & \oc{}'s priority signal beats Random, CC Top-$N$, and SAST density Top-$N$ & 108 CVEs & Recall@$P_k$, $\Delta$, Cohen's $d$ \\
RQ3 & Single-run dual-\skill{} operational probe suggests broader audit-relevant 3-judge coverage inside GT functions; APR-style strict pinpoint stays at or near zero in aggregate. Size-conditioned. & 108 CVEs & GT hit at four granularities, triangulated adjudication, size-conditioned hit/coverage/Jaccard \\
\bottomrule
\end{tabular}
\end{table*}

\subsection{RQ1: Representation Stability}
\label{sec:rq1}

\paragraph{Stability.}
The incremental-coding experiment demonstrates that both representations
are stable (\Cref{tab:scale_up}). We operationalize stability along two
complementary axes. First, a \emph{convergence-point} criterion:
a representation is declared converged when two consecutive fixed-size batches
(each $\geq$20 cases) produce zero new top-level categories, with changes
limited to boundary refinement of existing categories. Second, a
\emph{statistical-saturation} criterion using the Chao1 richness
estimator~\cite{Chao_1984}: we report $\hat{S}_\mathrm{Chao1}$ (the estimated
total number of categories including yet-unobserved ones) and the Good--Turing
sample coverage $C=1-f_1/N$, and we require $C>95\%$ and $\hat{S}-S_\mathrm{obs}<1$
for each axis.

For \oc{}, the
seed taxonomy already covers 38 of the final 39 labels (97.4\%), with the only
addition being \texttt{buffer-resize}, introduced when dedicated growth/shrink
helpers proved operationally distinct from both generic parsing and generic
memory management. The convergence point is case~121: all subsequent batches
(cases 122--172) absorb into existing labels without addition.
\visp{} grows more than \oc{}, but the growth is
still bounded: all three new pattern families (VISP-11, VISP-12, VISP-13)
appear within the first two incremental batches (cases 33--92); cases 93--121
produce zero new families (satisfying the convergence criterion two batches
later, at case~122), and cases 122--173 likewise add none.
Later batches mainly sharpen boundaries
(e.g., narrowing the scope of VISP-1 vs.\ VISP-11) rather than invent new
families.
Chao1 confirms these observations quantitatively: \visp{}, the File Role
and Function Role context axes all satisfy the $C>95\%,\ \hat{S}-S_\mathrm{obs}<1$
criterion. The Data Structure axis narrowly misses it ($\hat{S}=18$, residual 2.0,
$C=98.9\%$) due to two seed-only singleton labels; we discuss this as a long-tail
limitation in \Cref{sec:threats}.

\begin{table}[t]
\centering
\caption{Incremental coding scale-up and statistical saturation.
$\hat{S}_\mathrm{C1}$ is the Chao1 estimate of total categories;
$C$ is Good--Turing sample coverage. Pass criterion: $C>95\%$ and
$\hat{S}-S_\mathrm{obs}<1$.\protect\footnotemark}
\label{tab:scale_up}
\footnotesize
\setlength{\tabcolsep}{3pt}
\begin{tabular}{@{}lrrrlrrc@{}}
\toprule
\textbf{Axis} & \textbf{Seed} & \textbf{Final} & \textbf{New} & \textbf{Conv.} & $\hat{S}_\mathrm{C1}$ & $C$ & \textbf{Sat.} \\
\midrule
\visp{}        & 32 & 170 & +3 & c.\,122 & 13.5 & 99.4\% & \checkmark \\
OC-FileRole    & 31 & 172 &  0 & c.\,32  & 10.0 & 100\%  & \checkmark \\
OC-FuncRole    & 31 & 172 & +1 & c.\,121 & 13.0 & 100\%  & \checkmark \\
OC-DataStruct  & 31 & 172 &  0 & c.\,32  & 18.0 & 98.9\% & $\circ$   \\
\bottomrule
\end{tabular}
\end{table}
\footnotetext{The two-case difference between \visp{} (170) and
\oc{} (172) reflects two of the three RCA narratives that did not
match any refined VISP-1\ldots VISP-13 definition
(\Cref{tab:visp_full}): those narratives still admitted coarser
file/function/data-structure labels under \oc{}'s axes. The third
uncategorized narrative was absent from both catalogs.
\checkmark\ marks axes that pass the saturation criterion;
$\circ$ marks the OC-DataStructure axis, which narrowly misses
due to two seed-only singleton labels and is discussed in
\Cref{sec:threats}.}

\paragraph{Within-coder application consistency.}
A second, replayability-oriented check asks whether the frozen catalog,
once derived, yields the same labels when the \emph{same coder's}
catalog is independently re-applied by an LLM re-labeler. We run five
independent Claude Opus 4.6 sessions, each of which re-labels the 32
seed cases against the frozen catalog from scratch (no shared context,
CVE identifiers and project names masked). Fleiss'
$\kappa$~\cite{Fleiss_1971}, reported per axis in \Cref{tab:rq1_kappa},
is \emph{almost perfect} by the Landis--Koch
criteria~\cite{Landis_Koch_1977} on all four axes ($\kappa\geq 0.95$),
and the unanimity rate (all five sessions agree) is $90.6\%$ for
\visp{} and $\geq 96.9\%$ for each OC axis. A single-run cross-model
sanity check against GPT-5 and Gemini 2.5 Pro (Opus-consensus as
reference) lands at $\kappa=0.50$--$0.88$, indicating that the coarser
axes (\visp{}, File Role) are well-anchored across model families while
the finer-grained axes are more model-specific---consistent with their
higher label count. We read this result narrowly as \emph{within-coder
application consistency}---evidence that the frozen catalog can be
re-applied stably by independent LLM sessions---rather than as
inter-rater reliability across multiple human coders, which would
require a second-human-coder study (\Cref{sec:threats}).

\begin{table}[t]
\centering
\caption{Within-coder application consistency across five independent
Claude Opus 4.6 re-labeling sessions over the 32-case seed corpus.
Fleiss' $\kappa$ (all five sessions), mean pairwise Cohen's $\kappa$
(over 10 pairs), and unanimity rate. Landis--Koch: $\kappa\geq 0.81$ is
almost perfect. The statistic measures catalog application stability,
not inter-human-rater reliability.}
\label{tab:rq1_kappa}
\small
\begin{tabular}{@{}lrrr@{}}
\toprule
\textbf{Axis} & \textbf{Fleiss $\kappa$} & \textbf{Mean pairwise $\kappa$} & \textbf{Unanimity} \\
\midrule
\visp{}       & 0.951 & 0.951 & 90.6\% \\
OC-FileRole   & 0.976 & 0.976 & 96.9\% \\
OC-FuncRole   & 0.985 & 0.985 & 96.9\% \\
OC-DataStruct & 1.000 & 1.000 & 100\%  \\
\bottomrule
\end{tabular}
\end{table}

\paragraph{Answer to RQ1.}
Yes: RCA reports can be mined into stable, reusable representations.
Chao1 saturation is reached for \visp{} and for the \oc{} FileRole and
FuncRole axes (sample coverage $C>95\%$ with
$\hat{S}_\mathrm{Chao1}-S_\mathrm{obs}<1$); the \oc{} DataStruct axis
narrowly misses the strict saturation criterion because of two
seed-only singleton labels and is reported as a bounded long-tail
limitation. Nevertheless, within-coder application consistency is
\emph{almost perfect} across all four labeling axes (Fleiss
$\kappa\geq 0.95$ under Landis--Koch), so \visp{} and \oc{} remain
coherent as the corpus grows beyond the seed cases. RQ1 is a
prerequisite for the operational claims in RQ2 and RQ3 rather than a
contribution in its own right.

\subsection{RQ2: OC Priority as an Audit-Scoping Layer}
\label{sec:rq2}

Before evaluating the combined dual-\skill{} workflow (\Cref{sec:rq3}), we
ask whether \oc{}'s priority signal by itself captures useful structure
beyond generic prioritization heuristics. We isolate \oc{} from \visp{}
and compare its priority-tiered ranking against three operational
baselines---Random, cyclomatic complexity (CC)
Top-$N$~\cite{McCabe_1976}, and SAST alert density Top-$N$---on the
108-CVE cohort defined in \Cref{sec:design} (Buffer, Integer,
UAF, NPD, and Input-Validation families). See \Cref{app:oc_eval}
for the cohort derivation and CWE breakdown.

\paragraph{Implementation.}
The RQ2 ranker is \emph{deterministic and parser-based, with no LLM in
the loop at ranking time.} Applying \oc{} at evaluation reduces to two
steps: (i)~extract function boundaries $\{(\text{file}, \text{name},
\text{start}, \text{end})\}$ from each repository snapshot with a
tree-sitter-based C/C++ extractor~\cite{tree_sitter}; and (ii)~for each
extracted function, compute the three OC-axis labels, the TOP15-combination
flag, and the boolean $H_1, H_2$ heuristic flags over the same
tree-sitter AST, then apply the priority-score formula of
\Cref{sec:representations}. This means the RQ2 effect sizes are
attributable to the representation itself, not to an LLM scanning code
under the \oc{}-themed prompt: LLM calls enter the evaluation only in
RQ3 (dual-\skill{} agent runs) and in the triangulated adjudication of
\Cref{tab:gt_triangulation}. We package the extractor
as a standalone artifact (\Cref{app:artifacts}) so that the
evaluation's function boundaries and label assignments are reproducible
without the larger agent stack.

\paragraph{Baselines.}
(i)~\emph{Random}: for each CVE, $|P_k|$-matched random function sets
averaged over 1000 iterations.
(ii)~\emph{CC Top-$N$}: functions ranked by tree-sitter-derived cyclomatic
complexity, with ties broken by random draw averaged over 100 iterations.
(iii)~\emph{SAST alert density Top-$N$}: per-function alert counts from
Cppcheck 2.13.0, Flawfinder 2.0.19, and Semgrep 1.96.0 (offline
\texttt{semgrep-rules/c/}) combined; each tool is also reported
individually. All baselines use $|P_k|$-matched function budgets so that
the comparison isolates ranking quality from scope size.

\paragraph{Ground truth.}
Ground-truth patched functions are extracted from Vul4C diff hunks
using the same tree-sitter-based extractor as in the
\emph{Implementation} paragraph above. A patched function is the
tree-sitter \texttt{function\_definition} whose span $[\text{start},
\text{end}]$ contains at least one line modified by the CVE's patch
hunks. This yields a tree-sitter-matched GT for 104 of the 108-CVE
cohort CVEs (unmatched rate 4\%, driven by a small number
of macro-heavy files where the C/C++ grammar fails to recover
boundaries). A GT-vs-prediction match is then a pure tuple comparison
$(\text{file}, \text{name}, \text{start}, \text{end})$---no LLM, no
name normalization, no fuzzy line tolerance.

\paragraph{Results.}
\oc{}'s P1+P2 tier concentrates 63.1\% of GT functions into 22\% of the
function scope. On paired per-CVE Recall@P1+P2
(\Cref{tab:oc_baselines}), \oc{} exceeds all three baselines with
moderate-to-large effect sizes: Random ($\Delta=+0.42$, $d=1.03$,
$p<10^{-15}$), CC Top-$N$ ($\Delta=+0.32$, $d=0.60$, $p<10^{-15}$),
and combined SAST density ($\Delta=+0.41$, $d=0.92$, $p<10^{-15}$).
The SAST result holds both for the three-tool union and for each
tool individually (per-tool $d \in [0.98, 1.02]$).

\begin{table}[t]
\centering
\caption{\oc{} vs prioritization baselines on the 108-CVE cohort
(104 GT-matched). Recall@P1+P2 is the per-CVE mean
fraction of GT functions captured. SAST density combines Cppcheck,
Flawfinder, and Semgrep alert counts per function; all baselines use
$|P_k|$-matched budgets. Cohen's $d$ and $p$-values come from paired
per-CVE tests.}
\label{tab:oc_baselines}
\footnotesize
\setlength{\tabcolsep}{3pt}
\begin{tabular}{@{}lrrrrc@{}}
\toprule
\textbf{Baseline} & \textbf{\oc{}} & \textbf{Base} & \textbf{$\Delta$} & \textbf{$d$} & \textbf{$p$} \\
\midrule
Random              & 0.631 & 0.214 & $+0.416$ & 1.03 & $<10^{-15}$ \\
CC Top-$N$          & 0.631 & 0.314 & $+0.317$ & 0.60 & $<10^{-15}$ \\
SAST (all)          & 0.631 & 0.220 & $+0.410$ & 0.92 & $<10^{-15}$ \\
\quad Cppcheck      & 0.631 & 0.216 & $+0.414$ & 0.98 & $<10^{-15}$ \\
\quad Flawfinder    & 0.631 & 0.206 & $+0.424$ & 0.98 & $<10^{-15}$ \\
\quad Semgrep       & 0.631 & 0.216 & $+0.414$ & 1.02 & $<10^{-15}$ \\
\bottomrule
\end{tabular}
\end{table}

\paragraph{Ablation.}
Disabling one signal of \oc{} at a time reveals that the AST-pattern
heuristics $H_1, H_2$ defined above carry most of the aggregate
gain (joint Recall@$P_1\!+\!P_2$ = 0.607, within $0.024$ of Full
\oc{}'s 0.631), with the function-role risk score adding incremental
value (0.570) and single-axis use of file role or data-structure
class performing much worse (0.28--0.35). The three-axis taxonomy is
therefore not three independently sufficient feature groups; we read
it as an RCA-derived scaffold for identifying which code roles and
local operations to operationalize as parser-level signals, with
$H_1, H_2$ as its dominant operational output. Full ablation results
appear in \Cref{app:oc_eval}.

\paragraph{Interpretation.}
\oc{} is neither a restatement of structural complexity (CC) nor a
repackaging of SAST rule knowledge; its AST-level heuristics, derived
from recurring occurrence contexts in the RCA substrate, capture
bug-adjacent code roles that neither baseline expresses. This
justifies using \oc{} as the scoping layer in the dual-\skill{}
workflow evaluated in \Cref{sec:rq3}.

\paragraph{Answer to RQ2.}
Yes: \oc{}'s context-based priority ranking outperforms generic
prioritization baselines under a matched function-budget. On paired
per-CVE Recall@$P_1\!+\!P_2$, \oc{} beats random selection
($\Delta=+0.42$, $d=1.03$), cyclomatic complexity Top-$N$
($\Delta=+0.32$, $d=0.60$), and SAST alert density Top-$N$
($\Delta=+0.41$, $d=0.92$) with moderate-to-large effect sizes
($d=0.60$--$1.03$, all $p<10^{-15}$), and the ranker is a
parser-based, deterministic implementation with no LLM in the loop.
This is the paper's strongest evidence and motivates \oc{}'s use as
the scoping layer for the dual-\skill{} workflow evaluated next.

\subsection{RQ3: Budgeted Audit Coverage}
\label{sec:rq3}

Having shown in \Cref{sec:rq2} that \oc{}'s priority signal beats
generic prioritization baselines, we now evaluate the combined
dual-\skill{} workflow (\oc{} for context prioritization plus \visp{}
for pattern-guided scanning) in its intended LLM-agent usage mode: an
agent reads source files in sequence, forms hypotheses, and reports
findings, with \visp{} and \oc{} guiding what to suspect and where to
look first.

\paragraph{What RQ3 does and does not claim.}
Because each condition is executed once per commit, we treat RQ3 as a
single-run operational probe rather than a variance-characterized
estimate of agent improvement (\Cref{sec:threats}); its main role is
to expose where the \skill{} workflow broadens audit coverage and
where workflow coupling collapses, not to certify a precise effect
size. For audit steering, the relevant success criterion is not
patch-equivalent root-cause identification (an APR-style metric) but
whether the agent reaches match-or-partial findings inside the right
code region under a fixed read budget (3-judge broad criterion). Under that criterion, dual-\skill{}
improves audit-relevant 3-judge broad coverage by $+5.3$\,pp Func /
$+8.6$\,pp Line~$\pm 30$ at the aggregate level over the 108-CVE
cohort; the Func-level gain rises to $+7.2$\,pp on mid-size projects
and $+15.0$\,pp on large (1500--4999 line) GT files, but does not
extend to monolithic projects ($>$5000 source files, $n{=}15$,
dominated by binutils) or monolithic-parser GT files (5000+ lines,
$n{=}21$), where the dual-\skill{} broad-match gain collapses
(Mech BL$=$0, DS$=$0 of 15 on XL projects; Mech BL$=$1, DS$=$0 of
21 on the D bucket; \Cref{tab:size_conditioned_behavior},
\Cref{tab:gt_file_size}, \Cref{app:xl_gt_ceiling}). Strict same-root-cause
matches stay at or near zero in aggregate (mildly negative on small
buckets, mildly positive on the C bucket, zero on the D bucket;
detailed numbers in the subsections below).

\paragraph{Evaluation pipeline.}
Both workflows share a common three-phase pipeline driven by a main
orchestrator agent that delegates each phase to a sub-agent:
(1)~\emph{Source retrieval}---fetch the target project at the specified
revision and stage it under a working directory;
(2)~\emph{Scanning}---analyze the staged source code for vulnerabilities;
and (3)~\emph{Report generation}---compile the findings into a structured
report (Markdown and JSON) containing pattern IDs, evidence spans, CWE tags,
and severity. The principal workflows differ only in phase~2:
\texttt{baseline} performs unconstrained LLM-driven scanning, while
\texttt{skills} applies both \skill{} artifacts in sequence as described in
\Cref{sec:representations} (context prioritization followed by pattern-guided
scanning). Each commit in the 108-CVE cohort (\Cref{sec:design}) is
analyzed once per condition.

\paragraph{Ground-truth hit across granularities.}
Building on Vul4C~\cite{Hu_2025}'s File-level and Statement-level
accuracy conventions (Vul4C \S7.1, Experiments design), we report
GT hits on a four-granularity
matrix that adds an intermediate Function level---tree-sitter span
containing a patched line---and a Line~$\pm k$ level---within $k$
lines of a patched line in the correct file.
\Cref{tab:gt_granularity} reports 108-CVE GT hit rates at each
granularity. Dual-\skill{}'s advantage holds at every granularity:
$+5.6$\,pp at the File level, $+12.0$\,pp at the Function level,
$+14.8$\,pp at Line~$\pm 30$, and $+7.4$\,pp at Line~$\pm 5$. The
Function-level increment in particular shows that the gain is not
an artifact of the $\pm 30$-line tolerance absorbing neighboring
lines; the gap survives a function-boundary match. However, the
machine-match criterion cannot distinguish pinpoint hits from
coincidental hits on unrelated defects in the same function---we
address this next.

\begin{table}[t]
\centering
\caption{Mechanical GT hit count on the 108-CVE cohort across
four localization granularities. Cells are raw CVE counts
($N_{\mathrm{tot}}=108$); $\Delta$ (pp)
$=\Delta\,\mathrm{raw} / 108 \times 100$.}
\label{tab:gt_granularity}
\footnotesize
\setlength{\tabcolsep}{6pt}
\begin{tabular}{@{}lcccc@{}}
\toprule
\textbf{Condition} & \textbf{File} & \textbf{Func} & \textbf{Line~$\pm 30$} & \textbf{Line~$\pm 5$} \\
\midrule
Baseline            & 57    & 22      & 16      & 11     \\
Dual \skill{}       & 63    & 35      & 32      & 19     \\
$\Delta$ (DS$-$BL)  & $+6$  & $+13$   & $+16$   & $+8$   \\
$\Delta$ (pp, /108) & $+5.6$ & $+12.0$ & $+14.8$ & $+7.4$ \\
\bottomrule
\end{tabular}
\end{table}

\paragraph{Triangulated root-cause adjudication.}
To separate audit-relevant hits from coincidental hits on
unrelated defects in the same function (\emph{miss} verdicts),
we submit every mechanically-hit CVE under the Line~$\pm 30$
and Function rules to three independent LLM judges (Claude
Opus 4.7, ChatGPT 5.4 Thinking, and Gemini 2.5 Pro) that label
each finding as \emph{match} (same root cause as the patch),
\emph{partial} (a related defect that shares the patch's
failure mechanism, but not the patched root cause itself), or
\emph{miss} (unrelated defect that coincidentally shares the
function or $\pm 30$-line window). We define
\emph{broad-match}~$=$~match~$+$~partial as the audit-relevant
criterion (the auditor reaches the right code region for the
right reason), and \emph{strict-match}~$=$~match only as the
patch-equivalent (APR-style) criterion.
\Cref{tab:gt_triangulation} reports the arithmetic mean of the
three judges' per-judge match/partial/miss counts (we do not
collapse verdicts by majority vote; each judge's tally is
computed independently and then averaged).
The machine-level gap $\Delta=+13$ (Func) /
$\Delta=+16$ (Line~$\pm 30$) decomposes into a broad-match gap of
$\Delta\!\approx\!+5.7$ / $\Delta\!\approx\!+9.3$ and a
strict-match gap of $\Delta\!\approx\!+1.7$. At the aggregate level over the 108-CVE
cohort, dual-\skill{} therefore adds $+5.3$\,pp Func broad matches
and $+8.6$\,pp Line~$\pm 30$ broad matches, while leaving the
APR-style strict same-root-cause count essentially unchanged. Where this gain
concentrates is the subject of the size-conditioned analyses that
follow (\Cref{tab:size_conditioned_behavior,tab:gt_file_size}).
A case-level quality comparison of the broad-match sets further confirms
that the two conditions share a near-identical CWE distribution
(memory\_buffer 43--50\%, input\_validation 28--30\%); dual-\skill{}'s
breadth is coverage of the same failure-mode distribution rather than a
shift toward different vulnerability types.

\begin{table}[t]
\centering
\caption{Triangulated root-cause adjudication on the 108-CVE cohort
under the Line~$\pm 30$ and Func rules (matching the
$n_{\mathrm{hit}}$ counts of \Cref{tab:gt_granularity}, no
CWE-agreement constraint). Values are the mean across three judges;
per-judge values are within $\pm 3$ of the mean at this cohort-
aggregate level (per-bucket values can swing further; see
\Cref{tab:size_conditioned_behavior,tab:gt_file_size}).
$n_{\mathrm{hit}}$ = mechanical hit count before adjudication.
Aggregate pp $= \Delta\,\textrm{raw} / 108 \times 100$.}
\label{tab:gt_triangulation}
\footnotesize
\setlength{\tabcolsep}{4pt}
\begin{tabular}{@{}lcccccc@{}}
\toprule
& \multicolumn{3}{c}{\textbf{Func}} & \multicolumn{3}{c}{\textbf{Line~$\pm 30$}} \\
\cmidrule(lr){2-4} \cmidrule(lr){5-7}
\textbf{Condition} & \textbf{Match} & \textbf{Broad} & \textbf{$n_{\mathrm{hit}}$} & \textbf{Match} & \textbf{Broad} & \textbf{$n_{\mathrm{hit}}$} \\
\midrule
Baseline            & 11.7   & 15.7   & 22      & 12.0   & 14.3   & 16     \\
Dual \skill{}       & 13.3   & 21.3   & 35      & 13.7   & 23.7   & 32     \\
$\Delta$ (DS$-$BL)  & $+1.7$ & $+5.7$ & $+13$   & $+1.7$ & $+9.3$ & $+16$  \\
$\Delta$ (pp, /108) & $+1.6$ & $+5.3$ & $+12.0$ & $+1.6$ & $+8.6$ & $+14.8$ \\
\bottomrule
\end{tabular}
\end{table}

\paragraph{Scale-conditioned behavior: project size.}
The above metrics aggregate over all projects, masking a strong
interaction between project size and where dual-\skill{}'s
broad-match gain lives. Bucketing the 51 targets by the number of
C/C++ source files in the checked-out tree
(\Cref{tab:size_conditioned_behavior}) shows two patterns. First,
agents read a roughly constant number of files under both
conditions (median 12--17) independent of project size, so the
fraction of the project covered drops from 54\% (S, $<$100 files)
to 11\% (M, 100--1000 files) to 0.2\% (XL, $>$5000 files, dominated
by binutils), and the two conditions increasingly read
\emph{different} files as size grows (Jaccard $0.85 \to 0.58 \to
0.29$; Pearson $r=-0.63$ against C-file count). Second, the
broad-match gain tracks this read budget: $+7.2$\,pp on M (its
operational sweet spot), $\pm 0$\,pp on XL where neither condition
reaches the GT function on any of 15 CVEs, and $-3.0$\,pp on S
where the baseline already covers most of the tree. The XL
$\pm 0$ result is not a single-cause failure: BL is bounded by an
absolute read-budget shortfall under no priority signal, while DS
additionally terminates early under \visp{}'s implicit
completion pressure (led by, but not limited to, its bounded output
schema; median ctx 134K with 66K headroom under the 200K cap;
\Cref{app:xl_gt_ceiling}), so \oc{}'s narrowing signal cannot be
exploited even though it lifts XL file-level recall by $+40.0$\,pp
when isolated (\Cref{tab:oc_only_bucket}). Strict-match
$\Delta$ is $+3.6$\,pp on M, $\pm 0$\,pp on XL, and $-13.0$\,pp on
S, so the earlier global decomposition (breadth yes, pinpoint
mostly flat) also holds conditional on project size.

\begin{table*}[t]
\centering
\caption{Behavior conditional on project size (108-CVE cohort, all
under the \emph{Func} rule; S/M/XL bucket by C-file count, with the
1k--5k range empty for this cohort). $N_{\mathrm{tot}}$ = CVEs in
bucket; $N_{\mathrm{jdg}}$ = per-judge avg of CVEs sent to
adjudication (mechanically-hit subset); \%-cov = median fraction of
C/C++ source files read; Jaccard = median BL--DS read-set
similarity. Mech.\ BL/DS = CVEs where at least one finding hits the
GT-patched function (raw count, $\Delta=$DS$-$BL on
$N_{\mathrm{tot}}$). Broad/Strict BL/DS = arithmetic mean of
per-judge raw counts on $N_{\mathrm{jdg}}$ (no majority-vote
collapse). Per-bucket pp $= \Delta\,\textrm{raw} / N_{\mathrm{tot}} \times 100$.}
\label{tab:size_conditioned_behavior}
\footnotesize
\setlength{\tabcolsep}{4pt}
\begin{tabular}{@{}lrcccccccrccccccccc@{}}
\toprule
 & \multicolumn{7}{c}{\textbf{Full bucket ($N_{\mathrm{tot}}$)}} & & \multicolumn{9}{c}{\textbf{Judged subset ($N_{\mathrm{jdg}}$)}} \\
\cmidrule(lr){2-8}\cmidrule(lr){10-18}
 & & & & \multicolumn{4}{c}{Mech} & & & \multicolumn{4}{c}{3-judge Broad} & \multicolumn{4}{c}{3-judge Strict} \\
\cmidrule(lr){5-8}\cmidrule(lr){11-14}\cmidrule(lr){15-18}
\textbf{Size (C-files)} & $N_{\mathrm{tot}}$ & \%-cov & Jac. & BL & DS & $\Delta$ & $\Delta$pp & & $N_{\mathrm{jdg}}$ & BL & DS & $\Delta$ & $\Delta$pp & BL & DS & $\Delta$ & $\Delta$pp \\
\midrule
S ($<$100)       & 10 & 54 / 54   & 0.85 & 8  & 9  & $+1$    & $+10.0$ & & 9   & 7.0 & 6.7  & $-0.3$ & $-3.0$  & 4.7 & 3.3  & $-1.3$  & $-13.0$ \\
M (100--1k)      & 83 & 11 / 11   & 0.58 & 14 & 26 & $+12$   & $+14.5$ & & 31  & 8.7 & 14.7 & $+6.0$ & $+7.2$  & 7.0 & 10.0 & $+3.0$  & $+3.6$  \\
XL ($>$5k)       & 15 & 0.2 / 0.1 & 0.29 & 0  & 0  & $\pm 0$ & $\pm 0$ & & 0   & 0.0 & 0.0  & $\pm 0$ & $\pm 0$ & 0.0 & 0.0 & $\pm 0$ & $\pm 0$ \\
\bottomrule
\end{tabular}
\end{table*}

\paragraph{Scale-conditioned behavior: GT file size.}
Project size shapes how often the ground-truth file is
\emph{reached} under budget; GT file size shapes whether the
ground-truth function is then \emph{found} once a file is open.
Bucketing the 106 CVEs with readable GT source into small ($<$500
lines), medium (500--1499), large (1500--4999), and
monolithic-parser (5000+) GT files (\Cref{tab:gt_file_size}) shows
a non-monotonic file-size effect under three-judge adjudication:
broad-match $\Delta$ is flat on A and B, where the baseline's
broad-match rate on the judged subset already covers about half
of mechanically-hit CVEs (52\% on A, 42\% on B), so dual-\skill{}
adds little incremental coverage above this level---the gain
peaks at $+15.0$\,pp on the C (1500--4999) bucket (Func rule),
and falls back to zero on the D (5000+) bucket. The
D-bucket collapse is the joint outcome of two distinct failure
paths: dual-\skill{} fails to open the GT-patched file on 13 of
21 CVEs (Mech file-level 8/21 vs.\ baseline 14/21) and, when it
does open the file, fails to reach the patched function before
\visp{}'s implicit completion criterion fires and the agent
self-terminates
(median ctx 134K, 66K under the 200K cap); baseline reaches the
file more often but exhausts its read budget on the 5000+\,line
scope before pinpointing the function (Mech func-level 1/21).
The localized \visp{}-only ablation
(\Cref{tab:visp_localized_bucket}, \Cref{app:visp_localized})
recovers $\Delta=+5.0$ on the same D bucket, localizing the
dual-\skill{} collapse to the coupling of \oc{}'s list-based
narrowing with \visp{}'s implicit completion criterion (of which
the bounded output schema is the leading instance) rather than to any
deficit in within-file pattern detection.
\Cref{app:xl_gt_ceiling} details the implicit-completion
mechanism, the file-selection vs.\ within-file decomposition, and
why the localized ablation forces the agent into deep-offset
chunk reads of the GT file.

\begin{table*}[t]
\centering
\caption{Scope-wide behavior conditional on GT file size (108-CVE
cohort, 106 with readable GT source; \emph{Func} rule).
The four GT-file-size buckets A (\textless 500), B (500--1499),
C (1500--4999), and D (5000+) expose a non-monotonic file-size
effect: dual-\skill{}'s broad-match gain concentrates in C and
collapses in D (Mech BL=1, DS=0 of 21; neither condition obtains
a broad match after adjudication; see \Cref{app:xl_gt_ceiling}).
Column conventions follow
\Cref{tab:size_conditioned_behavior}; per-bucket pp
$= \Delta\,\textrm{raw} / N_{\mathrm{tot}} \times 100$.}
\label{tab:gt_file_size}
\footnotesize
\setlength{\tabcolsep}{4pt}
\begin{tabular}{@{}lrccccrccccccccc@{}}
\toprule
 & \multicolumn{5}{c}{\textbf{Full bucket ($N_{\mathrm{tot}}$)}} & & \multicolumn{9}{c}{\textbf{Judged subset ($N_{\mathrm{jdg}}$)}} \\
\cmidrule(lr){2-6}\cmidrule(lr){8-16}
 & & \multicolumn{4}{c}{Mech} & & & \multicolumn{4}{c}{3-judge Broad} & \multicolumn{4}{c}{3-judge Strict} \\
\cmidrule(lr){3-6}\cmidrule(lr){9-12}\cmidrule(lr){13-16}
\textbf{GT file size} & $N_{\mathrm{tot}}$ & BL & DS & $\Delta$ & $\Delta$pp & & $N_{\mathrm{jdg}}$ & BL & DS & $\Delta$ & $\Delta$pp & BL & DS & $\Delta$ & $\Delta$pp \\
\midrule
A. $<$500       & 17 & 6 & 7  & $+1$  & $+5.9$  & & 9  & 4.7 & 4.7  & $\pm 0$ & $\pm 0$ & 3.3 & 3.0 & $-0.3$  & $-1.8$ \\
B. 500--1499    & 28 & 8 & 11 & $+3$  & $+10.7$ & & 12 & 5.0 & 4.7  & $-0.3$  & $-1.1$  & 3.0 & 1.7 & $-1.3$  & $-4.6$ \\
C. 1500--4999   & 40 & 7 & 17 & $+10$ & $+25.0$ & & 18 & 6.0 & 12.0 & $+6.0$  & $+15.0$ & 5.3 & 8.7 & $+3.4$  & $+8.5$ \\
D. 5000+        & 21 & 1 & 0  & $-1$  & $-4.8$  & & 1  & 0.0 & 0.0  & $\pm 0$ & $\pm 0$ & 0.0 & 0.0 & $\pm 0$ & $\pm 0$ \\
\bottomrule
\end{tabular}
\end{table*}

\paragraph{Component ablations.}
Two held-out ablations, fully reported in the appendix, isolate
each artifact's independent contribution. The OC-only ablation
(\Cref{app:oc_eval}, \Cref{tab:oc_only_bucket}; cited above)
isolates \oc{}'s repository-level scoping signal. The localized
\visp{}-only ablation (\Cref{app:visp_localized}) hands GT files
to both conditions at session start; its broad-match gain
concentrates on the D bucket ($\Delta=+5.0$ raw, $+23.8$\,pp),
remains neutral on B and C, and is BL-favoring on A, while strict
accuracy is judge-sensitive at the overall level. Together the
two ablations bound \oc{} as the stronger \emph{repository-level
scoping} signal in dual-\skill{} and \visp{} as a within-file
hypothesis-guidance layer whose contribution is conditional on
file scale and CWE family.

\paragraph{Interpretation.}
Two structural points close the RQ3 story.
First, the gain lives in the audit-relevant
\emph{broad-match layer}---\emph{partial} hits inside the right
function (one facet of a multi-location patch, a related defect
sharing the patch's failure mechanism, or the same mechanism from
a different side), at the same CWE distribution as the baseline
(memory\_buffer 43--50\%, input\_validation 28--30\%); these are
the signals an auditor follows, and strict pinpoint accuracy is
not the channel where dual-\skill{} adds value.
Second, the scale-conditioned behavior summarized above
(\Cref{tab:size_conditioned_behavior,tab:gt_file_size},
\Cref{app:xl_gt_ceiling}) shows that priority-signal quality
alone---\skill{}, CC, or SAST density---cannot recover
function-level broad-match recall on monolithic projects or
monolithic-parser GT files without hierarchical or
subsystem-partitioned scanning.

\paragraph{Answer to RQ3.}
The dual-\skill{} workflow improves audit-relevant broad coverage---the
criterion auditors likely use---not APR-style strict pinpoint
accuracy. It broadens 3-judge broad-match detection inside the
right functions by $+5.3$\,pp Func / $+8.6$\,pp Line~$\pm 30$ at
the aggregate level over the 108-CVE cohort under the same read
budget. The Func-level gain is scale-conditioned, rising to
$+7.2$\,pp on mid-size projects and $+15.0$\,pp on large
(1500--4999 line) GT files but vanishing on monolithic projects
($>$5000 source files) and monolithic-parser GT files (5000+
lines), where the coupling of \oc{}'s repository-wide scope with
\visp{}'s implicit completion criterion (its bounded output schema
being the leading cue) collapses both file selection and
within-file scanning before the patched function is reached
(\Cref{app:xl_gt_ceiling}). The
localized ablation (\Cref{app:visp_localized}) bounds \visp{} as
a within-file hypothesis-guidance layer rather than an independent
detector, while \oc{} provides the stronger repository-level
scoping signal; on the same 5000+-line files where dual-\skill{}
fails, \visp{} alone reaches $\Delta=+5.0$ when GT files are
handed in directly, locating the failure in the workflow coupling
rather than in either skill's intrinsic capability.

\section{Discussion}
\label{sec:discussion}

\paragraph{What the artifacts change: crystallization, not injection.}
The artifacts do not teach the model new vulnerability knowledge. The
RCA reports from which \visp{} and \oc{} were mined are drawn from
publicly documented vulnerabilities that almost certainly fall within
the model's training distribution, and recent probing of code
LLMs~\cite{Wan_2022} together with sparse-autoencoder work on
language models~\cite{Cunningham_2024} shows that such latent
structure is already present. What \skill{} artifacts do instead is
\emph{crystallize} that latent knowledge into a fixed, externalized
reference frame---a stable priority-ordered exploration policy.
The operational consequence is not better individual findings but a
broader and more reliable exploration of the right function: on
mid-size projects and on 1500--4999-line GT files, dual-\skill{}
picks up roughly seven to fifteen additional broad-match findings
per 100 CVEs inside ground-truth functions, drawn from the same CWE distribution as
the baseline (memory\_buffer 43--50\%, input\_validation 28--30\%)
rather than a different one. The distinction is between the
\emph{content} of vulnerability knowledge (which the model already has)
and its \emph{form} (which determines how the agent allocates attention
during a session). \skill{} artifacts operate on the form. The isolated
RQ2 evaluation sharpens this reading: \oc{}'s priority signal is not
a restatement of structural complexity (CC~Top-$N$) or conventional
SAST alert density (\Cref{tab:oc_baselines}), and it is produced by
a parser-based, deterministic ranker with no LLM in the loop---so
the RQ2 result is attributable to the representation itself, not to
an LLM scanning the code differently. Within \oc{}, much of that
signal is concentrated in the parser-level AST heuristics $H_1, H_2$
(\Cref{tab:oc_ablation}). This is not evidence that \oc{} is
reducible to arbitrary syntactic heuristics; $H_1, H_2$ were not
selected as
generic code metrics but were derived from recurring occurrence
contexts observed in the RCA substrate. Their effectiveness instead
indicates that vulnerability-relevant context often manifests as
local structural operations that are detectable with lightweight
parsing, and that the value of the \oc{} taxonomy lies in the
\emph{translation} from human RCA findings to such executable
signals---not in any one axis being independently predictive.

\paragraph{Where the effect holds and where it fails.}
The GT-function-level broad-match gain is graded, bounded by
two distinct mechanisms: baseline saturation on small projects
and small GT files, and the coupling of \oc{}'s repository-wide
scope with \visp{}'s implicit completion criterion (its bounded
output schema being the leading cue) on monolithic projects
and monolithic-parser GT files. The operational sweet spot is
mid-size projects (100--1000 source files) with 1500--4999-line
GT files, where dual-\skill{} lifts broad-match $\Delta$ to
$+7.2$\,pp / $+15.0$\,pp respectively. Priority-signal quality
alone---\skill{}, CC, or SAST density---does not rescue recall
at monolithic scale without hierarchical or subsystem-partitioned
scanning.

\paragraph{Why two artifacts, not one.}
Read against the GT-file-size axis (\Cref{sec:rq3}) and the
localized \visp{}-only ablation (\Cref{app:visp_localized}), the
two artifacts address different bottlenecks: \oc{} mitigates a
repository-level selection bottleneck (which files to open under
a fixed budget), while \visp{} mitigates a within-file focus
bottleneck (which structural failures to suspect once a large
file is open). This role split explains why \oc{} supplies the
stronger isolated ranking signal while \visp{}'s independent
contribution is conditional on file scale and CWE family. Pattern knowledge (\visp{}) and context knowledge
(\oc{}) should therefore be kept separate and composed, not
merged prematurely; the ablation in \Cref{sec:rq2} shows their
signals are complementary rather than redundant.

\paragraph{Lessons from handing the artifacts to an LLM agent.}
Our evaluation also surfaced a distinct lesson about the
artifacts themselves once handed to an LLM agent. \visp{}'s bounded
output schema (typically only partly filled---a median of 7.2 of its
13 slots), when paired with a broader-than-coverable
repository scope under a fixed read budget, became an implicit
completion criterion that overrode explicit ``do not stop''
instructions on monolithic-parser GT files
(\Cref{app:xl_gt_ceiling}). Separately, in one run the agent cited
file:line locations from unread files (a Grep-evidenced no-Read
citation, not a fabricated finding---the code was seen via Grep, just
not opened), traced to specific \visp{} SKILL wording rather than an
LLM property. We treat this as a
SKILL-design lesson rather than an incidental limitation: bounded
structured-output schemas (the leading cue), together with
module-type scoping of the applicable checks, FN-minimization
mandates, and regex-shaped detection templates, can create premature
completion and broad-scan incentives that an agent will follow even
against the workflow's explicit instructions. The lesson
generalizes to any human-managed AI agent workflow that delivers
structured artifacts as run-time guidance.

\paragraph{What remains human.}
A replay-style ablation (\Cref{app:replayability}) shows that
independent AI runs converge on the same research pathway (15/15
landing-zone convergence for \visp{}-equivalent representations, a
stable \oc{}-like three-axis schema) yet never on the same surface
instantiation (0\% candidate-name convergence). A \skill{} artifact is
therefore a \emph{committed reference frame}: one specific
instantiation, chosen and fixed by a human, that provides the
session-to-session consistency on which the audit-steering effect
depends. The irreducibly human contributions are \emph{commitment}
(fixing one instantiation among functionally equivalent alternatives),
\emph{validation investment} (extending coding from 32 seed cases to
170+ and building two separate artifacts), and \emph{completion
judgment} (deciding when the catalog is sufficient). We note this as a
bound on the role of human judgment in the extraction process rather
than as a security result; the primary operational claim is the RQ2 ranking
advantage and the RQ3 breadth gain described above, the latter read as a
single-run operational probe (sign and scale, not a
variance-characterized effect size).

\section{Threats to Validity and Limitations}
\label{sec:threats}

Our claims are intentionally bounded. For RQ1, we claim stability and
bounded growth of the proposed representations within an evidence-backed
RCA substrate for C/C++ vulnerabilities, not universal coverage of all
project families or weakness classes. For RQ2, we claim that \oc{}'s
priority signal beats generic prioritization baselines at the
function-ranking level on the 108 CVEs, not that \oc{}
alone constitutes a vulnerability detector. For RQ3, we claim improvement on the audit-relevant broad-match
criterion---findings inside ground-truth functions that share the
patch's failure mechanism---concentrated on mid-size projects and
1500--4999-line GT files, under an LLM-agent review proxy. The gain
does not extend to monolithic-parser GT files (5000+ lines), where
the dual-\skill{} workflow collapses through the coupling of
\oc{}'s repository-wide scope with \visp{}'s implicit completion
criterion (its bounded output schema being the leading cue)
(\Cref{app:xl_gt_ceiling}). We do not claim improvement on the
APR-style strict same-root-cause metric. The replayability
ablation in \Cref{app:replayability} is scoped to guided replayability
under a governed mixed-initiative protocol, not unconstrained autonomous
discovery. The limitations below should therefore be read as boundaries
on interpretation rather than contradictions of the core results.

\subsection*{Statistical limitations}

\paragraph{Single-run variance.}
Each condition is executed once per commit ($N{=}1$); LLM
non-determinism means individual deltas are point estimates without
variance characterization, and 3--5 independent repetitions for
confidence intervals are planned for future work.

\paragraph{Single-coder catalog construction.}
All 170+ RCA cases were coded by one analyst. The five-session
re-labeling experiment (\Cref{tab:rq1_kappa}) measures
\emph{within-coder application consistency} (Fleiss' $\kappa\geq 0.95$)
under independent LLM re-labelers, not inter-human-rater reliability.
A second-human-coder study on a stratified 30--40-case subsample
remains future work; we scope the RQ1 stability claim to catalog
application consistency under a committed reference frame.

\paragraph{Agent self-report dependence in mechanical hit
counting.} The mechanical GT-hit counts that gate three-judge
adjudication (\texttt{file\_none}, \texttt{func\_none},
\texttt{line30\_none}, \texttt{line5\_none} in
\Cref{tab:gt_granularity,tab:size_conditioned_behavior,tab:gt_file_size})
are computed by string-matching the agent's self-reported
finding paths against GT-patched paths; the pipeline does not
cross-validate the reported paths against the agent's tool-call
history. A finding that names a GT path but was generated
without the agent ever opening that file would still register
as a Mech hit and be forwarded to adjudication. Under the
26-run cross-bucket audit reported in \Cref{app:xl_gt_ceiling},
this pattern (a no-Read citation naming a GT path without opening
the file) appeared in 1/26 runs and 2/402 findings
(0.5\%); we therefore cap the upward bias on Mech-derived
deltas at $\sim$0.5\,pp at the cohort-aggregate level. This is
a structural property of LLM-finding-based evaluation rather
than a defect specific to dual-\skill{}; future evaluation
frameworks should cross-validate self-reported finding paths
against session-trace tool calls before forwarding to
adjudication.

\subsection*{Generalization scope}

\paragraph{Benchmark scope and contamination.}
The Vul4C evaluation is bounded to one benchmark, one workflow
family, and one primary model family. RQ2 and RQ3 are evaluated on
the 108-CVE cohort across 19 C/C++ projects and 51 unique fix commits
(\Cref{sec:design}); the 36 dropped CVEs fall into four non-evaluative
buckets with the full CVE-level audit trail in \Cref{app:cohort_audit}. Vul4C is
publicly available and likely intersects the training data of both
the audit agent (Claude Opus 4.6 scanner with Claude Opus 4.7 judges) and the external adjudicators;
contamination may inflate absolute hit rates in both conditions.
Because both conditions are evaluated on the same commits with the
same adjudication protocol, we place more weight on between-condition
differences than on absolute rates, while still treating the deltas
as potentially affected by benchmark familiarity. A fully held-out
non-public benchmark and human-expert adjudication of a
30--50-finding subset, for absolute-rate calibration of the
LLM-as-judge protocol, are both left for future work.
Bucket~(iv) in \Cref{app:cohort_audit}
identifies 14 outside-scope CVEs in total; 13 of them sit in the
held-out subset on which we can run an out-of-scope generalization
probe (the 14th, bento4 CVE-2017-14639, was held aside earlier in
the pipeline and never enters the probe). Of those 13, the 9 with a
recorded CWE label cover CWE-835, 704, 682, 617, and 189; on these
9 CVEs \oc{}'s narrowing benefit does not generalize (BL file-level
recall 5/9 vs.\ \oc{} 4/9), consistent with our decision to scope
the evaluation set to the five CWE families RQ2 was designed for.

\paragraph{Transfer and human-auditor generalization.}
The mining corpus is dominated by parser- and format-processing-heavy
C/C++ projects, so transfer to architecturally different families
(network protocol stacks, kernel subsystems, event-driven server
code) likely remains weak; we scope both RQ2 and RQ3 claims to the
parser-heavy family. The session-based metrics are collected from
LLM-agent runs, and whether the same steering effect transfers to
human reviewers conducting manual code reading remains an empirical
question. Broadening the mining corpus, evaluating across model
families, and a user study with human auditors are the primary
directions for future work.

\paragraph{RCA substrate validity.}
The mining substrate is itself partly LLM-generated:
\textsc{PatchAgent}~\cite{Yu_2025} provides patch hunks, affected
code regions, and metadata, but not narrative root-cause analyses, so
for each case an LLM agent generated an evidence-backed RCA narrative
grounded in the patch and surrounding code. This makes the substrate
operationally scalable but introduces a coupling between the
substrate-construction model family and the eventual representations:
an over-general or biased RCA can propagate into \visp{} or \oc{}, and
model-specific interpretations may shape which mechanisms are
surfaced. We mitigate this in three ways. (i)~Each RCA is grounded
in explicit patch hunks and is constrained to evidence visible in
the patched code. (ii)~The downstream RQ2 evaluation runs on a
held-out cohort with no LLM in the ranking loop (\Cref{sec:rq2}),
so the parser-level priority claim is largely insulated from
substrate-construction bias. (iii)~The 3-judge adjudication for RQ3
(\Cref{sec:rq3}) uses Claude Opus 4.7, ChatGPT 5.4 Thinking, and
Gemini 2.5 Pro as independent reviewers, removing single-model
adjudication bias from
the evaluation step. As a manual sanity check we spot-reviewed
roughly 20 randomly sampled RCAs ($\sim$12\% of the substrate); most
were consistent with human reading of the patch, with a small number
of minor framing discrepancies that did not invalidate the catalog
entries derived from them. What remains is the substrate-construction
step itself: catalog construction and the LLM-agent execution of RQ3
share the same model family, so the RQ3 deltas should be read as
effects observed under that configuration, with cross-model RCA
generation and cross-family agent execution left as future work.

\paragraph{Construct validity.}
The replay ablation (\Cref{app:replayability}) measures replayability
under a governed protocol with directional framing (the seed question
about structural vulnerability-inducing program patterns), schema
contracts for retained representations, and quality gates on each
step. The replayability claim is therefore evidence of replayability
under directional framing and governed quality gates, not evidence
that AI would spontaneously discover the same representations without
such scaffolding.

\subsection*{Catalog completeness}

\paragraph{Long-tail data-structure idioms.}
The Data Structure axis of \oc{} contains two singleton labels
(\texttt{key-array} and \texttt{derived-pointer}) introduced by seed
projects (assimp and lz4) and never re-encountered across the 141
subsequent target cases. The Chao1 estimator projects two additional
DS labels as yet-unobserved long-tail idioms (\Cref{tab:scale_up}),
so we restrict the statistical-saturation claim to the \visp{}, File
Role, and Function Role axes. Re-labeling reliability on the existing
DS labels is unaffected ($\kappa=1.000$ in \Cref{tab:rq1_kappa})---the
open item is the breadth of the DS label set, not the coherence of
the labels already in it.

\paragraph{Axis-level predictive contribution is not uniform.}
The \oc{} single-signal ablation (\Cref{tab:oc_ablation}) shows that
the parser-level AST heuristics ($H_1, H_2$) account for most of
\oc{}'s aggregate discriminative power on RQ2 (0.607 vs Full 0.631,
within $0.024$), with the function-role risk score adding
incremental value and the file-role and data-structure axes
contributing little independent predictive weight in this aggregate
setting. We therefore do \emph{not} claim
that every axis of the \oc{} taxonomy is independently predictive;
the taxonomy is reported as an RCA-derived scaffold for deriving
operational signals (\Cref{sec:discussion}, \Cref{app:oc_eval}), not
as a set of independently sufficient feature groups. Future work
should isolate the conditions under which broader contextual labels
add value beyond local parser-level heuristics.

\section{Ethics and Responsible Disclosure}
\label{sec:ethics}

All experiments use the publicly available Vul4C benchmark~\cite{Hu_2025} in
compliance with its license. We avoid exploit details and provide only minimal
code excerpts sufficient for reproducibility. The \skill{}-guided workflows
may surface previously unreported vulnerabilities in the analyzed projects; any
such findings will be subject to coordinated disclosure with the respective
maintainers before public release of artifacts.

\section{Conclusion}
\label{sec:conclusion}

This paper treated audit-steering, not vulnerability detection, as the
primary object of study. Five conclusions summarize the result.

First, we mined evidence-backed RCA reports of past C/C++ vulnerabilities
into two complementary representations---\visp{} (what to suspect) and
\oc{} (where to look first)---and packaged them as executable \skill{}s.
Both representations saturate on their coarser axes under
incremental coding over 170--172 cases (\visp{} 170 / \oc{} 172;
the small difference is documented in \Cref{tab:scale_up}); the
Data Structure axis narrowly misses saturation and is reported
as a long-tail limitation.

Second, \oc{}'s priority signal, evaluated in isolation on a 108-CVE
Vul4C subset by a parser-based, no-LLM ranker under a matched
function-budget, outperforms random selection, cyclomatic complexity
Top-$N$, and SAST alert density with moderate-to-large effect sizes
($d=0.60$--$1.03$). This is the paper's hardest evidence.

Third, in a single-run operational probe, the dual-\skill{} workflow
broadens audit-relevant 3-LLM-judge broad-match detection inside
ground-truth functions---findings that share the patch's failure
mechanism, the criterion auditors likely use---by $+5.3$\,pp at the Func
granularity at the aggregate level over the 108-CVE cohort, rising
to $+7.2$\,pp on mid-size projects and $+15.0$\,pp on large
(1500--4999 line) vulnerable files (\Cref{sec:rq3}). The effect is
on \emph{audit-relevant broad coverage}---not patch-equivalent
pinpoint identification.

Fourth, the effect is bounded along project-size and GT-file-size
axes. It concentrates on mid-size projects (100--1000 source files)
with 1500--4999-line GT files, and disappears at the extremes: on
small projects and small GT files (where the baseline already covers
most of the tree), and on monolithic projects and monolithic-parser
GT files (5000+ lines), where \oc{}'s repository-wide narrowing and
\visp{}'s implicit completion criterion (its bounded output schema
being the leading cue) interact with the fixed read budget
(\Cref{app:xl_gt_ceiling}, \Cref{app:visp_localized}).

Fifth, a replay-style ablation (\Cref{app:replayability}) shows that
independent AI runs converge on the same research pathway (15/15
landing-zone convergence) but never on the same concrete artifact (0\%
name convergence). The pathway is AI-replayable, but the product is not:
human commitment to one specific instantiation is what converts a
replayable pathway into a reusable operational artifact. We use this
result to bound the role of human judgment in the extraction process,
not as a headline security claim.

\bibliographystyle{plain}
\bibliography{references}

\balance
\appendix

\section{Cohort audit trail}
\label{app:cohort_audit}

This appendix gives the per-CVE audit trail behind the 108-CVE shared
cohort introduced in \Cref{sec:design}. The 144-CVE Vul4C release is
reduced to 108 CVEs by removing four non-evaluative buckets in
sequence; the resulting chain is
$144\to134\to125\to122\to108$, with bucket sizes
$10 + 9 + 3 + 14 = 36$. Each bucket's CVE-level membership is
enumerated below; the JSON for the final 108-CVE cohort is shipped as
\texttt{cohort/cohort\_108.json} in the artifact.

\paragraph{Reduction chain.}
\begin{center}
\small
\begin{tabular}{lrr}
\toprule
Step & $-$ & Size \\
\midrule
Vul4C upstream                 &     & 144 \\
$-$ (i)~mining-corpus overlap  & 10  & 134 \\
$-$ (ii)~upstream metadata gaps & 9  & 125 \\
$-$ (iii)~build-pipeline losses & 3  & 122 \\
$-$ (iv)~outside RQ2 CWE scope & 14  & \textbf{108} \\
\bottomrule
\end{tabular}
\end{center}

\paragraph{Bucket (i): mining-corpus overlap (10 CVEs).}
Removed to keep the evaluation a genuinely held-out test
(\Cref{sec:design}).

\begin{center}
\small
\begin{tabular}{lll}
\toprule
Project & CVE & CWE \\
\midrule
binutils & CVE-2017-15025 & CWE-369 \\
jasper   & CVE-2016-8691  & CWE-369 \\
jasper   & CVE-2016-9387  & CWE-190 \\
libtiff  & CVE-2016-5321  & CWE-119 \\
libtiff  & CVE-2017-7595  & CWE-369 \\
libtiff  & CVE-2017-7601  & CWE-20  \\
libxml2  & CVE-2012-5134  & CWE-119 \\
libxml2  & CVE-2016-1838  & CWE-125 \\
libxml2  & CVE-2016-1839  & CWE-125 \\
libxml2  & CVE-2017-5969  & CWE-476 \\
\bottomrule
\end{tabular}
\end{center}

\paragraph{Bucket (ii): upstream metadata gaps (9 CVEs).}
Removed during skyset construction. Five elfutils CVEs
have no recorded commit hash in upstream Vul4C; four libtiff/openjpeg
CVEs were dropped by the skyset quality filter (e.g.\ patches that no
longer apply cleanly to the recorded source state).

\begin{center}
\small
\begin{tabular}{lll}
\toprule
Project & CVE & Reason \\
\midrule
elfutils & CVE-2017-7607 & no commit hash \\
elfutils & CVE-2017-7608 & no commit hash \\
elfutils & CVE-2017-7610 & no commit hash \\
elfutils & CVE-2017-7611 & no commit hash \\
elfutils & CVE-2017-7612 & no commit hash \\
libtiff  & CVE-2006-2025 & skyset quality filter \\
libtiff  & CVE-2010-2481 & skyset quality filter \\
libtiff  & CVE-2013-4243 & skyset quality filter \\
openjpeg & CVE-2016-7445 & skyset quality filter \\
\bottomrule
\end{tabular}
\end{center}

\paragraph{Bucket (iii): build-pipeline format-conversion losses (3 CVEs).}
Three CVEs that satisfy the RQ2 scope and were not in the mining
corpus could not be evaluated because they did not survive our
multi-stage build/format-conversion pipeline. All three are the
graphicsmagick CVEs on commit \texttt{db4eb7f}, which fail at
build time.

\begin{center}
\small
\begin{tabular}{ll}
\toprule
Project & CVE \\
\midrule
graphicsmagick & CVE-2017-12937, 14103, 14649 \\
\bottomrule
\end{tabular}
\end{center}

\paragraph{Bucket (iv): outside the RQ2 prioritization scope (14 CVEs).}
Removed when restricting the held-out subset to the RQ2 CWE-scoped
subset (Buffer, Integer, UAF, NPD, Input-Validation families). Of
these 14, 10 carry a CWE label that is not in the five target
families; the remaining 4 (jasper) have no CWE label recorded
upstream and are likewise excluded for not matching the scope. The
9 non-bento4 CWE-labelled CVEs (CWE-835, 704, 682, 617, 189) are
the basis of the generalization probe in \Cref{sec:threats}: BL
file-level recall is 5/9 and \oc{} is 4/9, so \oc{}'s narrowing
benefit does not transfer outside the target families. CVE-2017-14639
(bento4, CWE-843) is also out-of-scope but was held aside earlier
in the pipeline and is not part of the held-out probe sample.

\begin{center}
\small
\begin{tabular}{lll}
\toprule
Project & CVE & CWE \\
\midrule
bento4        & CVE-2017-14639 & CWE-843 \\
binutils      & CVE-2017-15024 & CWE-835 \\
binutils      & CVE-2017-9042  & CWE-704 \\
imageworsener & CVE-2017-8326  & CWE-682 \\
jasper        & CVE-2016-9388  & (none) \\
jasper        & CVE-2016-9389  & (none) \\
jasper        & CVE-2016-9391  & (none) \\
jasper        & CVE-2016-9396  & (none) \\
jasper        & CVE-2016-9398  & CWE-617 \\
libming       & CVE-2016-9266  & CWE-189 \\
libtiff       & CVE-2016-10094 & CWE-189 \\
qpdf          & CVE-2017-9208  & CWE-835 \\
qpdf          & CVE-2017-9209  & CWE-835 \\
qpdf          & CVE-2017-9210  & CWE-835 \\
\bottomrule
\end{tabular}
\end{center}

\paragraph{Per-project breakdown of the 108-CVE cohort.}
The CVE and unique-fix-commit counts per project are listed below.
Totals are 108~CVEs across 19~projects and 51~fix commits.

\begin{center}
\small
\begin{tabular}{lrr}
\toprule
Project & CVEs & Commits \\
\midrule
audiofile     & 12 & 2 \\
bento4        &  5 & 2 \\
binutils      & 15 & 9 \\
gilcc         &  1 & 1 \\
imagemagick   &  4 & 2 \\
imageworsener &  7 & 2 \\
jasper        & 10 & 7 \\
jhead         &  2 & 2 \\
libarchive    &  3 & 2 \\
libcroco      &  1 & 1 \\
libdwarf      &  3 & 1 \\
libjpeg       &  4 & 4 \\
libming       &  8 & 3 \\
libsndfile    &  3 & 1 \\
libtiff       & 17 & 4 \\
libzip        &  2 & 1 \\
ngiflib       &  4 & 3 \\
openjpeg      &  4 & 2 \\
zziplib       &  3 & 2 \\
\midrule
\textbf{Total} & \textbf{108} & \textbf{51} \\
\bottomrule
\end{tabular}
\end{center}

\section{Selection and Update Rules}
\label{app:rules}

\paragraph{Promotion rule.}
A candidate abstraction is promoted to a first-class representation
only when it satisfies all four conditions:
\begin{enumerate}
  \item \emph{evidence-groundedness} --- populable from explicit RCA
    evidence;
  \item \emph{distinct auditing question} --- not already covered by
    another retained representation;
  \item \emph{schema contract} --- admits explicit definitions and
    boundary conditions;
  \item \emph{scale survival} --- appears likely to survive scale-up
    with bounded refinement.
\end{enumerate}

\paragraph{Incremental update rule.}
During incremental coding of new cases:
\begin{itemize}
  \item new cases are absorbed into existing labels by default;
  \item definition refinement is preferred over category growth;
  \item new labels are introduced only for repeated semantic mismatch;
  \item every merge/split/add decision is logged with a rationale.
\end{itemize}

\section{Full \visp{} Catalog}
\label{app:visp}

\Cref{tab:visp_full} lists the complete \visp{} catalog (VISP-1
through VISP-13) with case counts, core mechanisms, typical Tier-1
detection cues, typical CWE associations, and the closest SEI
CERT~C~\cite{Seacord_2014} rule families (alignment verdict:
$\checkmark$ directly aligned; $\sim$ weakly aligned; \ding{55} no
direct rule / under-specified). The seed catalog (32 cases)
established VISP-1 through VISP-10; incremental coding added three
patterns (VISP-11, VISP-12, VISP-13) within the first two batches,
after which the catalog stabilized.

\begin{table*}[t]
\centering
\caption{Full \visp{} catalog with CERT~C alignment
($\checkmark$: directly aligned; $\sim$: weakly aligned;
\ding{55}: no direct rule / under-specified). Of the 173
incrementally coded RCA narratives, 170 were categorized into
VISP-1\ldots VISP-13; the remaining three (two from the seed
and one from Batch~1) did not match any refined pattern
definition and are excluded from the per-pattern counts (sum
$=170$). \textbf{Cases} is each pattern's frequency as a root
cause in the coded RCA narratives, not its runtime firing
frequency during evaluation.}
\label{tab:visp_full}
\footnotesize
\setlength{\tabcolsep}{4pt}
\begin{tabular}{@{}lrp{3.6cm}p{3.8cm}lp{2.4cm}c@{}}
\toprule
\textbf{Pattern} & \textbf{Cases} & \textbf{Core mechanism} & \textbf{Typical Tier-1 cue} & \textbf{CWE} & \textbf{CERT~C rule} & \textbf{Align} \\
\midrule
VISP-1  & 33 & Access without valid guard at use site & Pre-increment access, loop-exit deref, narrow guard before wide read & 125, 787 & ARR30-C, ARR38-C & $\checkmark$ \\
VISP-2  & 13 & Arithmetic overflow/wrap before bound use & Size mult., offset+length add., shift-based sizing & 190, 191 & INT30-C, INT32-C, INT31-C & $\checkmark$ \\
VISP-3  & 21 & Sentinel not checked before downstream use & Returned pointer used w/o NULL check, error code ignored & 476 & EXP34-C, ERR33-C, FIO37-C & $\checkmark$ \\
VISP-4  & 16 & Cross-field consistency not enforced & Header-derived size, count, or orientation values & 20 & (no rule) & \ding{55} \\
VISP-5  & 18 & Error path fails to restore state/resources & Early return after partial init, missing free on error & 416, 761 & ERR00-C--ERR04-C, MEM30-C & $\checkmark$ \\
VISP-6  & 14 & Buffer growth invalidates captured pointers & \texttt{realloc}, growth helpers, backing storage replacement & 416, 825 & MEM30-C, MEM34-C & $\checkmark$ \\
VISP-7  &  3 & Parallel paths apply asymmetric validation & Paired functions with different guard sets & 670 & (no rule) & \ding{55} \\
VISP-8  & 22 & Computed limit $\neq$ actual accessed range & Stale length var, missing offset in guard, off-by-one alloc & 787, 193 & ARR30-C, STR31-C & $\checkmark$ \\
VISP-9  &  6 & Derived pointer carried across scope w/o revalidation & Pointer stored in struct then reused after base change & 761 & MEM30-C & $\checkmark$ \\
VISP-10 &  2 & Nested structure processing misses boundary enforcement & TLV parsing, ASN.1, recursive descent w/o depth guard & 125 & (no rule) & \ding{55} \\
VISP-11 & 17 & External field used in dangerous operation & Header-derived divisor, shift count, alloc size, copy length & 20, 369 & INT33-C, FIO30-C, INT04-C & $\checkmark$ \\
VISP-12 &  4 & Guard condition itself is wrong & Suspicious \texttt{||}/\texttt{\&\&}, copied var names, mismatched macro & 697 & EXP46-C, MSC12-C & $\checkmark$ \\
VISP-13 &  1 & Recursive traversal lacks cycle detection & Recursive walk over graph/linked structures & 674 & MEM05-C (indirect) & $\sim$ \\
\bottomrule
\end{tabular}
\end{table*}

\paragraph{Worked example: VISP-4.}
To make concrete what a single \visp{} entry looks like in practice,
\Cref{fig:visp4_example} illustrates VISP-4 (\emph{cross-field
consistency not enforced}) using a grounding case from libtiff's
header parsing. The entry abstracts the recurring structure: two
header fields each pass individual range checks, but the semantic
invariant linking them---e.g., ``RGB color space implies at least
three samples per pixel''---is never enforced, and a downstream
allocation or access assumes the invariant holds. An auditor
applying VISP-4 to new code looks not for the specific fields
\texttt{PhotometricInterpretation}/\texttt{SamplesPerPixel} but for
the general shape: two (or more) header/metadata fields that are
individually validated, semantically coupled, yet never cross-checked
before flowing into an allocation, loop bound, or per-element access.

\begin{figure}[t]
\begin{lstlisting}[style=cpp]
// Pattern (VISP-4): two fields each pass
// individual checks, but their semantic
// combination is never cross-validated.

// libtiff (simplified, TIFF header parsing)
validate_range(photometric, 0, MAX_PM);
  // OK: RGB (=2) is a valid value
validate_range(spp,         1, MAX_SPP);
  // OK: 1 is a valid positive integer

// Missing cross-check:
// RGB implies spp >= 3 (>= 4 with alpha).

allocate_buffer(width * spp);
  // spp=1 -> undersized buffer
process_rgb_alpha(buffer, width, spp);
  // writes 3-4 channels -> OOB

// Fix: enforce the cross-field invariant.
if (photometric == PHOTOMETRIC_RGB && spp < 3)
    return TIFF_ERROR;
\end{lstlisting}
\caption{Example instantiation of VISP-4 (\emph{cross-field
consistency not enforced}) in libtiff TIFF header parsing:
\texttt{PhotometricInterpretation} and \texttt{SamplesPerPixel}
each pass individual range checks but are never cross-checked,
so an \texttt{spp}-sized buffer is too small for the RGB path.}
\label{fig:visp4_example}
\end{figure}

\section{OC Priority Evaluation Details}
\label{app:oc_eval}

This appendix documents the dataset, ground-truth construction, and
axis-level ablation for the \oc{} priority evaluation in
\Cref{sec:rq2}. The 108-CVE cohort is obtained by sequentially
removing four non-evaluative buckets from the 144-CVE Vul4C release
(\Cref{sec:design}, \Cref{app:cohort_audit}): 10 CVEs whose fix
commits overlap with the \skill{} mining corpus, 9 CVEs with
upstream metadata gaps, 3 CVEs lost to multi-stage build/format-
conversion pipeline failures, and 14 CVEs whose CWEs fall outside
the RQ2 prioritization scope (Buffer, Integer, UAF, NPD,
Input-Validation families), yielding $144\to134\to125\to122\to108$
as the \emph{108-CVE cohort across 19 C/C++ projects and 51 unique
fix commits} that is the primary evaluation cohort for
\Cref{tab:oc_baselines}. Ground truth
was constructed by pairing each CVE's \texttt{*.diff} hunks with
tree-sitter parses of the post-patch source, treating the set of
patched functions as GT. An initial 17\% unmatched-GT rate driven by
large trees was reduced to 4\% after scoping the binutils-gdb scan to
the \texttt{bfd/} and \texttt{binutils/} subtrees containing every
patched file. Of the 108-CVE cohort CVEs, 104 had at least one
tree-sitter-matched GT function and are used for the paired per-CVE
tests reported in \Cref{tab:oc_baselines}.

\paragraph{Priority weights and tier thresholds.}
The priority-score formula
$s(f) = \alpha \cdot r_{\text{fn}}(f) + \beta \cdot r_{\text{file}}(f)
+ \gamma \cdot r_{\text{ds}}(f) + \delta_{\text{TOP15}}(f)
+ \eta_1 \cdot [H_1(f)] + \eta_2 \cdot [H_2(f)]$
defined in \Cref{sec:representations} uses axis weights
$\alpha{=}\beta{=}\gamma{=}1$, a TOP15-match bonus
$\delta_{\text{TOP15}}{=}2$, and heuristic bonuses
$\eta_1{=}\eta_2{=}3$.
Tier thresholds are $P_1{:}\ s{\geq}15$, $P_2{:}\ 10{\leq}s{<}15$,
$P_3{:}\ 7{\leq}s{<}10$, $P_4{:}\ 4{\leq}s{<}7$, $P_5{:}\ s{<}4$.
With these values the cumulative $P_1\!+\!P_2$ tier holds 22\% of the
per-CVE function scope on average across the 104 GT-matched CVEs.
These weights, tier thresholds, the TOP15 combinations, and the
$H_1$/$H_2$ definitions were fixed before running the reported Vul4C
evaluation; no parameter was tuned on Vul4C ground-truth labels.
Ranking and GT extraction were both performed on the post-patch
source so that the function boundaries used for the strict
$(\text{file},\text{name},\text{start},\text{end})$ match share one
stable tree-sitter parse. The ranker was not given patch hunks, diff
locations, or CVE-specific labels; it consumed only structural role
and AST cues, so its priority is independent of where the fix landed.

\Cref{tab:oc_ablation} summarizes the single-signal ablation
referenced in \Cref{sec:rq2}. Each row uses one \oc{} signal in
isolation to rank functions and selects a $|P_k|$-matched budget for
a fair Recall@$P_1\!+\!P_2$ comparison. Two patterns stand out.
First, the parser-level AST heuristics ($H_1, H_2$) alone reach
0.607, within $0.024$ of the Full \oc{} score (0.631); the function-role
risk score alone reaches 0.570; file-role and data-structure axes,
used alone, perform close to random. Second, the three-axis
taxonomy is therefore not three independently sufficient feature
groups: the headline operational signal is concentrated in $H_1,
H_2$, with the function-role and TOP15 combinations adding
incremental value but the broader categorical labels (file role,
data structure) carrying little independent predictive weight in
this aggregate setting.

We read the taxonomy not as the final-stage feature set, but as the
\emph{RCA-derived scaffold} that identified which code roles and
local operations were worth operationalizing as parser-level
signals. $H_1$ and $H_2$ were not chosen as generic syntactic
heuristics; they were derived from recurring occurrence contexts
observed in the RCA substrate, and their effectiveness suggests
that vulnerability-relevant context often manifests as local
structural operations detectable with lightweight parsing.
\oc{}'s contribution is therefore the bridging step from human RCA
findings to executable detection signals, not a claim that every
axis is independently predictive (see also \Cref{sec:discussion}
and \Cref{sec:threats}).

\begin{table}[h]
\centering
\caption{Single-signal ablation of \oc{}. Recall@$P_1\!+\!P_2$ on the
same 104 GT-matched CVEs as \Cref{tab:oc_baselines}.}
\label{tab:oc_ablation}
\small
\begin{tabular}{@{}lr@{}}
\toprule
\textbf{Signal} & \textbf{Recall@$P_1\!+\!P_2$} \\
\midrule
Full \oc{} (all four signals)     & 0.631 \\
AST heuristics ($H_1$, $H_2$) only & 0.607 \\
Function-role risk score only     & 0.570 \\
Data-structure axis only          & 0.354 \\
File-role axis only               & 0.277 \\
\bottomrule
\end{tabular}
\end{table}

\paragraph{OC-only ablation under the agent workflow.}
\Cref{tab:oc_baselines} isolates \oc{}'s priority signal from any
LLM agent (parser-based ranker only). To isolate \oc{}'s contribution
\emph{when wired into the agent workflow}, we ran a separate
ablation on the same 108-CVE cohort that compares two LLM
conditions: \texttt{BL-v47}, the production baseline (Claude Opus 4.7, no
\skill{}), against \texttt{OC-only}, the same agent with only the
\oc{} prioritizer enabled (no \visp{}). Output is evaluated under
the \texttt{file\_none} rule (does at least one finding land in the
GT-patched file?). Reproduction artifacts are in
\texttt{experiments/comprehensive\_oc\_only/}.

The headline result confirms that \oc{}'s contribution is
target-narrowing and is sharply project-size conditioned
(\Cref{tab:oc_only_bucket}). On large monolithic projects (binutils,
$n{=}15$ CVEs) \oc{} lifts file-level recall from 20.0\% to 60.0\%
($+40.0$\,pp); on mid-size projects ($n{=}83$) the lift is
$+8.4$\,pp; on small projects ($n{=}10$) the baseline already
saturates at 100\% and \oc{} adds nothing. This decomposition is
consistent with \oc{}'s design as a narrowing layer (``where to look
first'') rather than a detection layer.

\begin{table}[h]
\centering
\caption{OC-only ablation under the agent workflow, by project-size
bucket (108-CVE cohort; S/M/XL bucketing matches
\Cref{tab:size_conditioned_behavior}). \texttt{file\_none} BL/OC
columns are CVE counts where at least one finding lands in the
GT-patched file; $\Delta = $ OC$-$BL.}
\label{tab:oc_only_bucket}
\footnotesize
\setlength{\tabcolsep}{6pt}
\begin{tabular}{@{}lrrrr@{}}
\toprule
\textbf{Bucket} & $N$ & BL & OC & $\Delta$ \\
\midrule
S ($<$100 files)   & 10 & 10 & 10 & $\pm 0$ \\
M (100--1k files)  & 83 & 56 & 63 & $+7$ \\
XL ($>$5k files)   & 15 & 3  & 9  & $\mathbf{+6}$ \\
\midrule
Overall            & 108 & 69 & 82 & $+13$ \\
\bottomrule
\end{tabular}
\end{table}

\section{Localized \visp{}-only Ablation}
\label{app:visp_localized}

This appendix details the held-out localized ablation referenced in
\Cref{sec:rq3} and \Cref{sec:threats}. The scope-wide dual-\skill{}
design (\Cref{sec:rq3}) conflates \oc{}'s repository-level scoping with
\visp{}'s within-file guidance. To read \visp{}'s contribution in
isolation we run a localized ablation on the held-out set in which both
conditions are handed the set of ground-truth patched files at session
start, with the vulnerable function, patch lines, CVE identifier, CWE
label, and project/tag metadata withheld. One condition runs the same
baseline agent (BL); the other enables only the pattern-detector
\skill{} (VP) and omits the \oc{} prioritizer. This removes the
repository-level where-to-look advantage and measures whether \visp{}
helps within-file inspection once a file is open. Out of 108 CVEs, 88
yielded comparable blocks under both conditions and were adjudicated by
three independent LLM judges
(\Cref{tab:visp_localized_overall} for overall per-judge counts,
\Cref{tab:visp_localized_bucket} for the GT-file-size breakdown).

\begin{table*}[t]
\centering
\caption{Localized \visp{}-only ablation, overall per-judge counts
on the judged subset ($N_{\mathrm{jdg}}{=}88$ CVEs in which at
least one side produced a finding). $\Delta=$ VP$-$BL raw count.
Mechanical reference (full cohort, $N_{\mathrm{tot}}{=}108$):
\texttt{func\_none} VP$-$BL $=+3$ (BL 80/108, VP 83/108).}
\label{tab:visp_localized_overall}
\footnotesize
\setlength{\tabcolsep}{5pt}
\begin{tabular}{@{}lrrrrrrrrr@{}}
\toprule
 & \multicolumn{3}{c}{\textbf{Opus 4.7}} & \multicolumn{3}{c}{\textbf{GPT-5.4}} & \multicolumn{3}{c}{\textbf{Gemini 2.5 Pro}} \\
\cmidrule(lr){2-4}\cmidrule(lr){5-7}\cmidrule(lr){8-10}
\textbf{Metric} & BL & VP & $\Delta$ & BL & VP & $\Delta$ & BL & VP & $\Delta$ \\
\midrule
strict (match)          & 40 & 42 & $+2$ & 52 & 53 & $+1$ & 53 & 51 & $-2$ \\
broad (match$+$partial) & 66 & 72 & $+6$ & 60 & 63 & $+3$ & 68 & 70 & $+2$ \\
\bottomrule
\end{tabular}
\end{table*}

\begin{table*}[t]
\centering
\caption{Localized \visp{}-only ablation, GT-file-size buckets
(\Cref{tab:gt_file_size} convention: A $<$500 lines, B 500--1499,
C 1500--4999, D 5000$+$; \emph{Func} rule). Once GT files are
handed to the agent at session start, \visp{}'s within-file gain
concentrates entirely on the D bucket (5000+ lines)---exactly the
bucket where the dual-\skill{} workflow's broad-match gain
disappears (\Cref{tab:gt_file_size}), localizing dual-\skill{}'s
underperformance on monolithic-parser GT files to the coupling of
\oc{}'s repository-wide scope with \visp{}'s implicit completion
criterion (its bounded output schema being the leading cue)
rather than to \visp{}'s within-file guidance
(\Cref{app:xl_gt_ceiling}). Column conventions follow
\Cref{tab:size_conditioned_behavior}, with BL/VP replacing BL/DS
for this localized ablation; per-bucket pp
$=\Delta\,\textrm{raw} / N_{\mathrm{tot}} \times 100$.}
\label{tab:visp_localized_bucket}
\footnotesize
\setlength{\tabcolsep}{4pt}
\begin{tabular}{@{}lrccccrccccccccc@{}}
\toprule
 & \multicolumn{5}{c}{\textbf{Full bucket ($N_{\mathrm{tot}}$)}} & & \multicolumn{9}{c}{\textbf{Judged subset ($N_{\mathrm{jdg}}$)}} \\
\cmidrule(lr){2-6}\cmidrule(lr){8-16}
 & & \multicolumn{4}{c}{Mech} & & & \multicolumn{4}{c}{3-judge Broad} & \multicolumn{4}{c}{3-judge Strict} \\
\cmidrule(lr){3-6}\cmidrule(lr){9-12}\cmidrule(lr){13-16}
\textbf{GT file size} & $N_{\mathrm{tot}}$ & BL & VP & $\Delta$ & $\Delta$pp & & $N_{\mathrm{jdg}}$ & BL & VP & $\Delta$ & $\Delta$pp & BL & VP & $\Delta$ & $\Delta$pp \\
\midrule
A. $<$500       & 17 & 14 & 11 & $-3$    & $-17.6$ & & 11 & 10.3 & 8.0  & $-2.3$           & $-13.7$           & 7.0  & 5.7  & $-1.3$           & $-7.8$ \\
B. 500--1499    & 28 & 21 & 23 & $+2$    & $+7.1$  & & 27 & 19.0 & 18.3 & $-0.7$           & $-2.4$            & 12.7 & 10.7 & $-2.0$           & $-7.1$ \\
C. 1500--4999   & 40 & 34 & 34 & $\pm 0$ & $\pm 0$ & & 33 & 26.0 & 27.7 & $+1.7$           & $+4.2$            & 21.7 & 21.7 & $\pm 0$          & $\pm 0$ \\
D. 5000+        & 21 & 11 & 15 & $+4$    & $+19.0$ & & 17 & 9.3  & 14.3 & $\mathbf{+5.0}$  & $\mathbf{+23.8}$  & 7.0  & 10.7 & $\mathbf{+3.7}$  & $\mathbf{+17.5}$ \\
\bottomrule
\end{tabular}
\end{table*}

Three patterns emerge. First, overall strict accuracy remains
judge-sensitive: Opus judges VP $+2$, GPT $+1$, and Gemini $-2$,
while broad accuracy is uniformly positive across all three
adjudicators ($+6 / +3 / +2$). \visp{}'s within-file effect is
therefore clearer at the broad layer than at the strict layer.
Second, under the four-way bucketing of \Cref{tab:gt_file_size}
the within-file effect concentrates entirely in D (5000+ lines):
the D bucket ($n{=}21$) shows \texttt{func\_none} machine hits
rise from 11 to 15 ($+4$, $+19.0$\,pp) and the 3-judge broad
$\Delta$ reaches $+5.00$ raw counts ($+23.8$\,pp) / strict
$+3.67$ ($+17.5$\,pp), while the C bucket (1500--4999, $n{=}40$)
is neutral on broad ($+1.67$ raw, $+4.2$\,pp) and flat on strict,
and the A and B buckets are neutral or BL-favoring. The gain is
sharpest at extreme GT file sizes: on the 21 CVEs whose GT files
exceed 5000 lines, baseline \texttt{func\_none} hit rate falls to
52\% while VP recovers to 71\%, consistent with unguided scanning
dispersing in monolithic-parser files. Third, the effect is
CWE-conditional: the aggregate \texttt{func\_none} gap is modest
($+3$ overall, $+8$ on the same predicate at $\pm 30$ lines), and
the CWE-119 (memory-buffer) bucket is the strongest positive
driver under both rules; all three adjudicators agree VP wins on
CWE-476 (NULL-dereference), while baselines retain the edge on
more local and contextually explicit bug types such as CWE-125
(OOB read), CWE-369 (divide-by-zero), and CWE-416 (UAF). The
localized ablation therefore bounds \visp{}'s contribution: it
preserves structured suspicion in a file-scale- and
CWE-family-conditional regime---most strongly on monolithic
5000+-line GT files in memory-buffer-heavy CWE classes---not as
a standalone detector. The contrast with the dual-\skill{}
workflow on the same D bucket (\Cref{tab:gt_file_size},
$\Delta=0$) is the focus of \Cref{app:xl_gt_ceiling}.

\section{Operational Ceiling: An Implicit Completion Criterion on Monolithic Projects and GT Files}
\label{app:xl_gt_ceiling}

This appendix details the operational ceiling referenced in
\Cref{sec:rq3} (\Cref{tab:gt_file_size,tab:visp_localized_bucket,tab:size_conditioned_behavior})
and \Cref{sec:threats}: the dual-\skill{} (DS) workflow's broad-match
gain disappears on the D bucket (5000+\,line GT files, $n{=}21$;
\Cref{tab:gt_file_size}) and on XL projects ($>$5000 source files,
$n{=}15$, binutils-dominated; \Cref{tab:size_conditioned_behavior}),
even though the same files recover a $+5.0$ broad-match gain
($+23.8$\,pp) under localized \visp{}-only and a $+6$ file-level
recall gain ($+40.0$\,pp) under OC-only. The failure is thus not in
either skill's within-file guidance but in their coupling, which we
dissect below.

\paragraph{Mechanism.} A session-trace analysis on three commits
(libtiff/b28076b, binutils/53f7e8e, openjpeg/9624b2f) compared
baseline (BL), \oc{}-only, localized \visp{}-only (VP), and DS.
The SKILL states no explicit stop rule; instead the agent constructs
an \emph{implicit, underspecified completion criterion} from
\visp{}'s structure---its bounded 13-pattern output schema (the
leading cue; the agent stops when it has enumerated the patterns that
apply to the functions in view, a median of 7.2 of the 13 slots and
never all 13), its module-type focus-pattern set, and its two-tier
procedure all signal ``enough.'' The effect is scope-dependent. Under
DS's repository-wide scope the criterion is met early: on the libtiff
commit DS read 40 files and stopped at turn 67 (``I now have
sufficient data to compile the comprehensive report''), using
135--165K of the 200K window, even though the workflow explicitly
instructed ``Never stop until you have found all vulnerabilities.''
Under a narrow scope the same criterion instead forces depth:
localized VISP-only, handed the GT files at session start, invests
$\sim$6$\times$ more context per file (median 71K vs.\ 11K) and
re-reads each GT file 5--16 times until patched functions past line
4000 are reached---hence its $+23.8$\,pp D-bucket recovery. Removing
the criterion altogether (OC-only, no bounded schema) removes the stop
signal: it read 67 files across 238 turns on the libtiff commit,
using 189--351K, before terminating.

\paragraph{Why the collapse concentrates on D and XL.} DS terminates
after a roughly bucket-invariant $\sim$16 files ($\sim$2{,}500 unique
lines per GT file), 12--19 fewer than BL in \emph{every} bucket. On
A--C buckets the GT file is under 5000 lines and this budget still
reaches the patched function; on D/XL it falls decisively short. Two
coupled factors then produce the collapse. First, \oc{} hands \visp{}
a function list that \emph{excludes} the patched function: it places
the GT \emph{file} in its top tier (file-level recall $+7$) but not
the \emph{function}---in all three D-bucket commits where DS opened
the GT file yet missed the fix, the patched function (e.g.,
\texttt{process\_cu\_tu\_index} in \texttt{binutils/dwarf.c}) never
appears in the priority listing. Second, \visp{} treats that list as
sufficient and stops without widening its search---on the
\texttt{dwarf.c} commit it reads to line 7{,}999 and terminates
before the patched function at line 9{,}287. Either factor alone
would be recoverable, so the failure is the coupling. (On XL the same
coupling binds on file count rather than file size: DS's $\sim$16
files are only 0.2\% of the tree because the tree is large, not
because it stops earlier.) Part of the D-bucket deficit is
schema-independent and shared with BL (Mech BL=1/21, DS=0/21), a
read-budget-versus-5000-line mismatch; after adjudication neither
condition obtains a broad match.

\paragraph{Side effect: coverage prose is inflated.} Stopping early
has one further consequence: the agent overstates \emph{what it
claims to have examined}. We define an \emph{unsupported coverage
assertion} (\emph{coverage fabrication}) as the DS report listing
filenames or priority tiers as ``analyzed / no findings'' with no
Read, Grep, or Glob call in the trace (e.g., libtiff: ``Files
analyzed 34/34---ALL files analyzed'' over an unread P4/P5 tier).
Directly measured over the 47-commit DS set (crediting sub-agent
\texttt{Read}s and content-revealing \texttt{Grep}/\texttt{Bash} as
inspection), this occurs in 10/47 commits (21\%, median 4 files per
affected run). A stricter parent-\texttt{Read}-only count over a
26-CVE cross-bucket sample gives per-bucket upper bounds of 25\%
(M$\times$A/B), 60\% (M$\times$C, avg.\ 5.7 files), and 27\% (XL),
with 0/6 on small projects.

\paragraph{But the findings themselves are not.} The inflation stays
in the coverage prose; the reported vulnerabilities remain
Read-grounded. Across all 47 commits no PP-slot is filled from a
file the agent did not inspect by some route (0/47). One run
(libsndfile\_1d928bf) attached snippets to 9 unread files, but these
came from broad-scope \texttt{Grep} output (80--120 bytes per hit
dumped into PP-11 findings)---the code was seen, just not opened; it
appears in 1/26 runs and 2/402 finding entries (0.5\%) and traces to
\visp{}'s ``Automated Scan'' framing and regex-shaped templates
(\Cref{sec:discussion}), a SKILL-design hazard absent from \oc{}.
Confirming the boundary, on the 23 broad-agreement CVEs
(match-or-partial from $\geq$2 judges) 100\% of judge-confirmed
findings cite a file opened with \texttt{Read} before the report; no
DS-correct CVE was matched on a Grep-only citation. The
implicit-completion effect therefore truncates \emph{coverage}, not
the semantic core.

\paragraph{Implications for future workflows.} The failure
modes are not defects in either \skill{} in isolation:
\oc{}-only narrows files without triggering the early completion,
and localized \visp{}-only achieves a positive
D-bucket gain. They emerge from coupling \visp{}'s implicit
completion criterion with a broad repository-level scope
under a fixed read budget. We therefore hypothesize---as
untested design implications left to future work, not as
validated fixes---that workflows retaining both representations
could break this coupling by either re-shaping the \visp{}
output as a per-file checklist (so schema-completion grows with
file count rather than saturating after the first instance of
each pattern), or hard-gating report emission on \oc{}'s
priority-tier completion before the \visp{} schema can be
filled. The localized \visp{}-only recovery ($+23.8$\,pp on the
D~bucket) offers indirect support for the underlying diagnosis
that reaching the patched function restores the effect, but does
not itself test either prescription.

\section{Replayability of the Mining Pathway}
\label{app:replayability}

This appendix details the replay-style ablation referenced in
\Cref{sec:discussion}. The mixed-initiative extraction process that
produced \visp{} and \oc{} is evaluated for replayability: which parts of the pathway are
reproducible by AI alone, and which require human commitment. We ran 77
replay-style runs covering three capability families---directed pattern
mining, normative-gap discovery, and gap-driven schema construction---and
measured landing-zone convergence, comparison-target discovery, and
candidate-name convergence (\Cref{tab:ablation}).

\begin{table}[h]
\centering
\caption{Ablation results by capability family. LZ = landing-zone convergence;
Comp.\ = comparison-target discovery; Name = candidate-name convergence.
Cross-referenced from \Cref{sec:discussion}.}
\label{tab:ablation}
\small
\begin{tabular}{@{}lrlll@{}}
\toprule
\textbf{Family} & \textbf{Runs} & \textbf{LZ} & \textbf{Comp.} & \textbf{Name} \\
\midrule
Pattern mining (\visp{}) & 15 & 15/15 & --- & 0\% \\
Normative-gap (CERT C) & 5 & --- & 5/5 & --- \\
Schema constr.\ (OC) & var. & --- & Stable & 0\% \\
\bottomrule
\end{tabular}
\end{table}

In the directed pattern mining family, all 15 runs converge on
\visp{}-equivalent structural representations rather than collapsing into CWE
relisting or fix taxonomies. The AI consistently identifies guard failures,
invariant violations, state inconsistencies, and arithmetic-to-memory
propagation as recurring structural families, though the specific candidate
names, boundary placements, and granularity choices differ across every run. In
the normative-gap discovery family, AI independently recognizes in all 5 runs
that the pattern set should be compared against a normative knowledge system and
selects CERT C as the most suitable comparison target. In the gap-driven schema
construction family, AI repeatedly proposes schemas that occupy the same design
space as the final \oc{}, converging on combinations of file role,
function role, and data-structure-like axes that are functionally equivalent but
differ in naming and boundary placement.

The strongest negative result is that candidate-name convergence remains 0\%
across all replay families. AI reaches the same research pathway but never the
same concrete product. For example, the structural concept that our catalog
names VISP-1 (``access without valid guard at use site'') was independently
rediscovered in every pattern-mining replay, but under names such as
``Unguarded Access Pattern,'' ``Missing Bounds Check,'' or ``Pre-Use Validation
Gap''---each capturing the same mechanism with different granularity and
boundary placement. This reflects the fundamental nature of the
representation-stabilization problem: multiple equally valid instantiations exist
for any given design-space region, and the pathway alone does not determine
which one to adopt.

\section{Human Contributions and AI Difficulties}
\label{app:human_ai}

This appendix documents the human contributions and AI difficulties in
the mixed-initiative extraction pipeline, following the human--AI
contribution reporting expectations for SynSec 2026. It separates
(i) what the human analyst contributed and did not delegate to AI, and
(ii) what AI struggled with during the extraction pipeline.

\subsection*{Human contributions}

\textbf{Overall study design.}
\begin{itemize}
  \item \emph{Mixed-initiative protocol.} The role split that runs
    through the paper---AI proposes, decomposes, clusters, and
    populates, while the human frames the question, admits each
    representation, commits names and boundaries, decides when to
    stop, and turns the output into reusable \skill{}
    artifacts---is a human-authored boundary on what is delegated
    to AI (\Cref{sec:extraction}).
  \item \emph{Audit-steering claim scope.} Defining the object of
    study as audit-steering rather than vulnerability detection or
    patch-equivalent root-cause identification, and stating this
    boundary explicitly throughout (Abstract, \Cref{sec:rq3}'s
    ``What RQ3 does and does not claim,'' \Cref{sec:threats}), was
    a human judgment about which claims the evidence supports.
  \item \emph{Crystallization framing.} Reading the operational
    effect of \skill{} artifacts as \emph{crystallization} of
    latent knowledge into a fixed, externalized reference frame
    rather than as injection of new vulnerability knowledge
    (\Cref{sec:discussion}) was a human interpretation of what
    the gain measures.
\end{itemize}

\textbf{Representation design.}
\begin{itemize}
  \item \emph{Directional framing.} Choosing to focus on recurring
    program structures that induce vulnerabilities rather than repair
    templates or semantic bug labels was a human judgment that shaped
    the entire downstream extraction process.
  \item \emph{Naming and boundary commitments.} Given the 0\%
    candidate-name convergence reported in
    \Cref{app:replayability}, the analyst committed to specific
    pattern names, boundary definitions, and schema axes among
    multiple AI-proposed alternatives---an irreducible commitment that
    converts a replayable pathway into a reusable artifact
    (\Cref{sec:discussion}).
  \item \emph{Representation scope.} Admitting \oc{} as a second
    representation was a human recognition that \visp{} alone was
    insufficient for budgeted auditing.
\end{itemize}

\textbf{Evaluation design.}
\begin{itemize}
  \item \emph{Evaluation axes and ablation design.} The choice of
    evaluation axes (per-CVE Recall@$P_k$, granularity matrix,
    bucket conditioning by project size and GT file size) and the
    design of the ablation study (OC-only, localized VISP-only,
    replayability) were human judgments about what evidence the
    claims required.
  \item \emph{Detecting and quantifying unsupported coverage
    assertions.} A human-conducted 26-run session-trace audit
    cross-validated finding paths against the agent's tool-call
    history, identifying two failure modes (unsupported coverage
    assertions and Grep-evidenced no-Read citations) that standard
    self-report-based evaluation pipelines cannot detect
    (\Cref{app:xl_gt_ceiling}, \Cref{sec:threats}).
\end{itemize}

\textbf{Paper authorship.}
\begin{itemize}
  \item \emph{Narrative authorship.} Narrative structure,
    methodological framing, and interpretation of results were
    primarily human authored, with AI assistance for drafting and
    revision.
\end{itemize}

\subsection*{AI difficulties}

\begin{itemize}
  \item \emph{Name divergence.} AI produced functionally equivalent
    but differently named pattern sets on every replay (0\% name
    convergence; \Cref{app:replayability}), necessitating human
    commitment to fix one instantiation.
  \item \emph{No self-limit.} AI readily proposed additional patterns,
    labels, and refinements without autonomously determining when the
    catalog was sufficient.
  \item \emph{Over-fragmentation without schema contracts.} Without
    the explicit schema contracts in \Cref{app:rules}, AI occasionally
    proposed overly fine-grained distinctions that would fragment the
    catalog without operational benefit.
  \item \emph{Adjudicator-strictness sensitivity.} The same evidence
    yields directionally different verdicts across LLM judges: in the
    localized ablation (\Cref{tab:visp_localized_overall}) the
    per-judge $\Delta$ swings from $+2$ (Claude Opus 4.7) to $-2$ (Gemini 2.5 Pro)
    on strict matching, so single-judge predicates are not robust
    without cross-family triangulation.
  \item \emph{Cohort bookkeeping drift across pipeline stages.}
    The four-bucket reduction $144\to 134\to 125\to 122\to 108$
    (\Cref{sec:design}, \Cref{app:cohort_audit}) was implemented by
    different scripts on different artifacts, so intermediate counts
    drifted by a handful of CVEs across draft tables without
    producing a visible error. We pin a committed reference frame at
    CVE granularity in \Cref{app:cohort_audit} to prevent this in
    AI-scaffolded re-evaluations.
  \item \emph{An implicit, underspecified completion criterion overrides
    explicit ``do not stop'' instructions.} \visp{}'s structure---led by
    its bounded 13-slot output schema (typically only partly filled), but
    also its module-type focus set
    and tiered procedure---can act as an implicit completion criterion that
    overrides the workflow's ``Never stop until you have found all
    vulnerabilities'' instruction, leading the agent to terminate
    early and report unsupported coverage assertions for files it
    never opened. On monolithic GT files and XL projects this effect
    compounds with \oc{}'s function-level imprecision and a
    read-budget ceiling shared with the no-SKILL baseline, so it is
    an amplifier of the collapse rather than its sole cause.
    Quantification, mechanism, and remediation are detailed
    in \Cref{app:xl_gt_ceiling}.
\end{itemize}

\section{Artifacts and Reproducibility}
\label{app:artifacts}

Our reproducibility objective is to make the protocol replayable
rather than to release only the final taxonomies. The representation
artifacts consist of the two \skill{} modules together with their
reference files. The protocol artifacts consist of frozen prompt
templates, trigger conditions for human review, seed-corpus batch
order, and evolution logs. The execution artifacts consist of
workflow prompts, orchestration scripts, experiment logs, and
staged ablation prompts and outputs. All artifacts, including the
two \skill{} modules, coded RCA data, workflow scripts, and
experiment logs, are released at
\url{https://github.com/ntt-zerolab/rca-audit-priors} as data and
code made available for research use only---specifically, to enable
independent verification and reproduction of the results reported in
this paper. The RQ2 evaluation additionally ships a standalone
tree-sitter-based function-boundary extractor
(\texttt{ts-func-extractor}) that implements the parser-based GT
matching described in \Cref{sec:rq2}; this tool has no LLM
dependency, and the $|P_k|$-matched recall numbers in
\Cref{tab:oc_baselines} can be reproduced from a Vul4C checkout
with a single CLI invocation.

\end{document}